\pdfoutput=1

\documentclass[iop]{emulateapj}
\usepackage{amsmath,url,natbib,graphicx}
\usepackage[breaklinks,colorlinks,citecolor=blue]{hyperref}
\usepackage[all]{hypcap}
\newcommand{\ang}{\,\text{\AA}}

\newcommand{\hr}{\,{\rm hr}}

\newcommand{\um}{\,{\rm \mu m}}
\newcommand{\us}{\,{\rm \mu s}}

\newcommand{\inch}{\arcsec}
\newcommand{\seconds}{\,\mathrm{s}}
\newcommand{\mK}{\,{\rm mK}}
\newcommand{\Gb}{\,{\rm Gb}}
\newcommand{\ev}{\,{\rm eV}}

\shorttitle{The ARCONS Pipeline}
\shortauthors{van~Eyken et al.}
\begin{document}

\title{The ARCONS Pipeline: Data Reduction for MKID Arrays}

\author{J.~C.~van Eyken,\altaffilmark{1,2,3}
  M.~J.~Strader,\altaffilmark{1}
  A.~B.~Walter,\altaffilmark{1}
  S.~R.~Meeker,\altaffilmark{1}
  P.~Szypryt,\altaffilmark{1}
  C. Stoughton,\altaffilmark{4}
  K.~O'Brien,\altaffilmark{5}
  D.~Marsden,\altaffilmark{1} 
  N.~K.~Rice,\altaffilmark{1}
  Y.~Lin,\altaffilmark{1}
  B.~A.~Mazin\altaffilmark{1}
  }

  \email{vaneyken@ipac.caltech.edu}

  \altaffiltext{1}{Department of
    Physics, UC Santa Barbara, Santa Barbara, CA 93106, USA}
  \altaffiltext{2}{Las Cumbres Observatory Global Telescope Network,
    6740 Cortona Dr. suite 102, Goleta, CA 93117, USA}
  \altaffiltext{3}{Now at NASA Exoplanet Science Institute, California
    Institute of Technology, 770 South Wilson Avenue, M/S 100-22,
    Pasadena, CA 91125, USA}
  \altaffiltext{4}{Fermilab Center for
    Particle Astrophysics, Batavia, IL 60510, USA}
  \altaffiltext{5}{Department of Physics, University of Oxford, Denys
    Wilkinson Building, Keble Road, Oxford, OX1 3RH, UK}

\begin{abstract}
The \textbf{A}rray \textbf{C}amera for \textbf{O}ptical to
\textbf{N}ear-IR \textbf{S}pectrophotometry, or ARCONS, is a camera
based on Microwave Kinetic Inductance Detectors (MKIDs), a new
technology that has the potential for broad application in
astronomy. Using an array of MKIDs, the instrument is able to produce
time-resolved imaging and low-resolution spectroscopy constructed from
detections of individual photons. The arrival time and energy of each
photon are recorded in a manner similar to X-ray calorimetry, but at
higher photon fluxes. The technique works over a very large wavelength
range, is free from fundamental read noise and dark-current
limitations, and provides microsecond-level timing resolution. Since
the instrument reads out all pixels continuously while exposing, there
is no loss of active exposure time to readout. The technology requires
a different approach to data reduction compared to conventional
CCDs. We outline here the prototype data reduction pipeline developed
for ARCONS, though many of the principles are also more broadly applicable to
energy-resolved photon counting arrays (e.g., transition edge sensors,
superconducting tunnel junctions). We describe the pipeline's current
status, and the algorithms and techniques employed in taking data from
the arrival of photons at the MKID array to the production of images,
spectra, and time-resolved light curves.

\end{abstract}


%------------------------------------------------------------------
\section{Introduction}
The recent development of superconducting Microwave Kinetic Inductance
Detectors \citep[MKIDs;][]{Day2003, Mazin2012} opens the door to a
promising new generation of UV-to-near-IR astronomical imaging
instruments.  By constructing an array of MKIDs it is possible to
build a detector analogous to a conventional CCD that is additionally
able to resolve individual photons, recording their arrival times and
energies in a manner similar to X-ray microcalorimeters. It is
therefore possible to obtain time-resolved imaging and spectroscopy in
a single observation. The wavelength range can be very large,
potentially as much as $\approx 0.1 - 5\um$ (in comparison
to $\approx 0.3 - 1\um$ for CCDs), with a theoretical energy resolution limit as
high as $R=E/\Delta E = \lambda/\Delta\lambda \sim 100$ for a $100\mK$
operating temperature, and with microsecond time resolution. The
devices do not suffer from the intrinsic readout-noise and
dark-current issues inherent to traditional CCDs, and as the pixels
can be read out simultaneously and continuously, there is no loss of
on-sky exposure time during readout.  Data can be read and processed
in real-time, potentially allowing for live science-image telescope
guiding and flux monitoring with no loss of throughput, as well as
focal-plane speckle control for high-contrast imaging
\citep[][S.~R.~Meeker et al., 2015 in preparation]{Martinache2014}.

The \textbf{Ar}ray \textbf{C}amera for \textbf{O}ptical to
\textbf{N}ear-IR \textbf{S}pectrophotometry \citep[ARCONS;
see][]{ARCONS} is a prototype MKID instrument which has now been
tested in the field at both the Lick $120\inch$ and Palomar $200\inch$
telescopes. The current Palomar design consists of a 2024 pixel
($44\times 46$) MKID array, with a median spectral resolution
$R\approx 6.5$ at $4000\ang$, and approximately 70\% useable
pixel yield. The instrument is sensitive across a bandwidth of
$\approx 4000-11000\ang$, constrained only by the optics
passband. Since the technology is still in its infancy, we expect to
make significant improvement in the specifications with
time. Nevertheless, the first trial telescope runs have already
yielded preliminary science results, including the suggestion of
optical enhancements in early giant pulses from the Crab Pulsar
\citep{Strader2013}, and the detection of orbital expansion in the AM
CVn binary system, SDSS J0926+3624 \citep{Szypryt2014}.

A detailed description of ARCONS is given by \citet{ARCONS}. Here we
describe the prototype data reduction process developed to handle the
ARCONS data. The current development version of the software can be
found online at \url{https://github.com/bmazin/ARCONS-pipeline}. The
data reduction pipeline is a work in progress, and much refinement is
expected as the MKID technology matures; the basic components are
in place, however, to produce images, spectroscopy, and timing analyses.

Section \ref{sec:outline} provides a brief overview of the main
sections of the pipeline and describes the main data products.
Section \ref{sec:readout} describes the firmware algorithms used to
produce the raw data output from the camera. Section
\ref{sec:pipeline} describes each of the steps in the data reduction
and calibration in detail, from the initial photon detection through to fully
calibrated photon event lists, and in the process touches on some 
characterization of the current ARCONS instrument. Section
\ref{sec:outputproducts} describes the process of obtaining three main
science data products: images, spectroscopy, and photometry. A brief
summary follows in Section \ref{sec:conclusions}.

\section{Outline}\label{sec:outline}

The ARCONS pipeline is written in Python, and makes extensive use of
the fast array handling and vectorization capabilities provided by the
NumPy and SciPy packages \citep{NumPy-SciPy},\footnote{Eric Jones,
  Travis Oliphant, Pearu Peterson and others. SciPy: Open Source
  Scientific Tools for Python, 2001--. \url{http://www.scipy.org/}}
along with astronomical tools from the Astropy package
\citep{astropy},\footnote{\url{http://www.astropy.org}} and the data
visualization tools from the Matplotlib package
\citep{Hunter2007}. Owing to the large quantities of data to be
handled, some care has been taken to optimize the various processing
algorithms for speed and efficient disk access, though there remains
room for further optimization. The PyTables library\footnote{Francesc
  Alted, Ivan Vilata, and others, PyTables: Hierarchical Datasets in
  Python, 2002--. \url{http://www.pytables.org}} provides an interface
to the Hierarchical Data Format (HDF5) standard for file
handling,\footnote{The HDF Group. Hierarchical data format version 5,
  2000--2010. \url{http://www.hdfgroup.org/HDF5/}.}  which is the
primary data storage format used in the pipeline. The HDF5 standard
allows for versatile structuring of very large data sets and provides
various tools for efficient manipulation of on-disk data.

The full reduction process can be divided into three broad steps: the
creation of the raw data read from the camera; reduction of the data
to produce fully calibrated and flagged photon-event lists; and the
creation of the final data output products (images, spectra, or lightcurves)
from those photon lists.

The continuous monitoring and readout of each pixel occurs largely in
firmware (Section \ref{sec:readout}). The raw data are saved in
HDF5-format, as are the various intermediate data products created
later in the pipeline. To keep file sizes manageable, raw data are
written to a new file every five minutes, yielding typical file sizes
of a few hundred Mb to $\sim 1\Gb$ for our instrument configuration
(depending on the total flux observed). We refer to these as
``observation files.'' Various metadata are also recorded in the
files, including the beammap used to associate each MKID resonator
with the correct pixel location in the array (see Section
\ref{sec:beammapping}), telescope pointing and status information, and
the precise start time of the data in the file.

A number of steps are performed to calibrate the data. Cosmic rays can
be cleaned (Section \ref{sec:cosmicrays}), dead or unassigned pixels
are masked, and intermittent ``hot'' pixels are identified (Section
\ref{sec:timemasks}).  Exposures of laser reference sources are used
to calibrate the mapping of resonator phase-shift to wavelength for
each photon (Section \ref{sec:wavecal}). Twilight sky exposures are
used to normalize the responses of the pixels with respect to each
other as a function of wavelength
(Section \ref{sec:flatcal}). An overall spectral response calibration is then
applied by observing a known spectral standard, in order to
flatten the effective wavelength response of the instrument (Section
\ref{sec:speccal}). Where necessary, nonlinearity due to high count
rates is accounted for (Section \ref{sec:linearity}). Finally, astrometric
calibrations are performed to identify the precise location of each
detector pixel on the sky at a given time, and hence assign sky
coordinates to each detected photon (Section
\ref{sec:astrometry}). Most of these steps involve the creation of a
small intermediate calibration file that is used to apply
corrections to one or more raw observation files.

The result of this processing is a calibrated photon list for every
raw data file (Section
\ref{sec:photlist}). The associated per-pixel exposure time mask is
also stored in the same file, as well as a copy of all calibration files applied, in
order to keep an easily traceable audit trail for diagnostic purposes.

With the photon lists in hand, we can produce several science products. For
imaging, the photon lists can be read into a module which stacks
the photons onto a virtual pixel grid, using the exposure time masks to weight
the photons according to the effective total exposure time in each virtual
pixel (Section \ref{sec:imaging}). The wavelength range can optionally be
restricted to create images in different bands, which, if desired, can then be
combined to create color images.

Alternatively, time-resolved photometry can be performed similarly to
traditional CCD photometry (Section \ref{sec:photometry}). If the
effective integration times required are short, this can be performed
on the calibrated data without running it first through the full image
stacking module. If longer integration times are desired, it is
possible to run photometry on the stacked and de-rotated
images. Again, the photometry can be restricted in wavelength to
emulate photometric filtering, allowing for simultaneous multi-band
photometry where desired. Where high timing precision is required,
photon arrival times can be corrected to Solar System barycentric
time.

If spectra are required (Section \ref{sec:spectroscopy}), the photons
for a chosen aperture can similarly be extracted, with or without
prior image stacking, and the energy plotted as a function of
wavelength integrated over a given effective exposure time (analogous
to Integral Field Spectroscopy).

The ultimate data product (currently partially
implemented) will be a four-dimensional data ``hypercube,''
essentially a four-dimensional histogram of photon events with
dimensions of RA, dec, wavelength, and time. We describe each of the
steps in the current pipeline in more detail below.

\section{ARCONS Firmware: Creating the Raw Data}\label{sec:readout}

\subsection{Pulse Detection}\label{sec:pulsedetection}

\begin{figure} \epsscale{1.12}
\epsscale{1.12}
\centering
\plotone{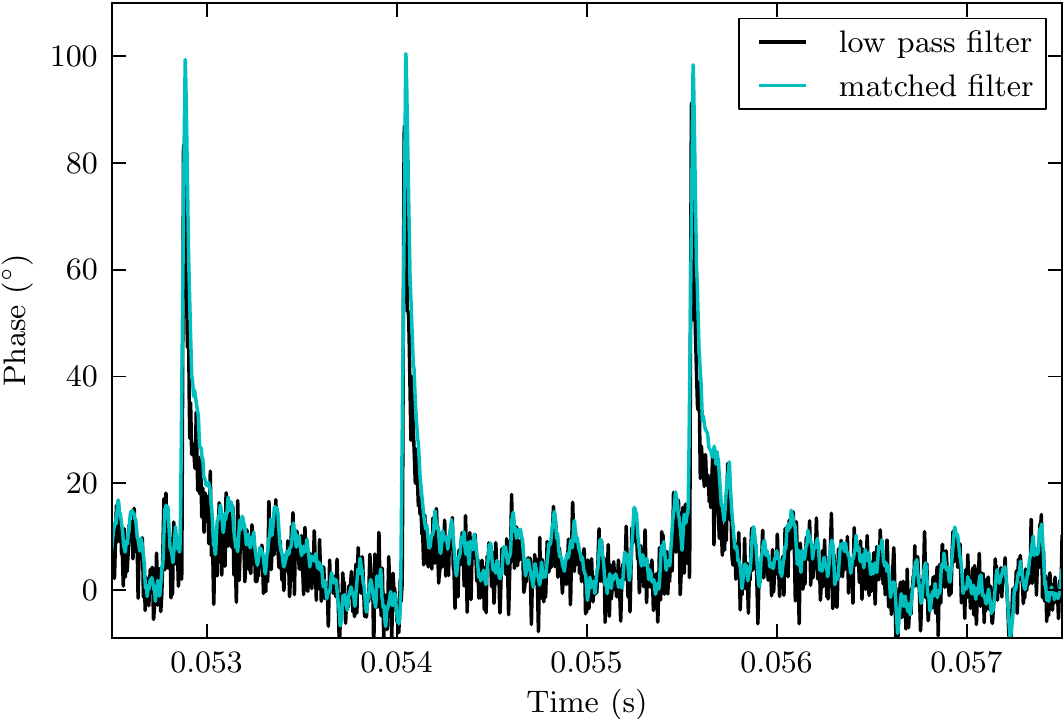}
\caption{Three photon pulses as seen in the phase timestream from one MKID,
where the phase has passed through a low-pass filter with a corner frequency at
250 kHz (black line) or a matched filter made with an average pulse
template (cyan line).}
\label{fig:photonPulses}
\end{figure}

The firmware for the digital readout for ARCONS is described in detail
by \citet{McHugh2012}.  Each pixel comprises a small superconducting
resonant circuit (a single MKID) tuned to a unique resonant
frequency. A microlens array with a 92\% fill factor focuses
  the light onto the inductor part of the pixel.  When a photon is
  absorbed in the inductor, the inductance of the circuit, and
therefore its resonant frequency, briefly changes. Each resonator is
driven by a microwave probe tone, the phase of which will
change with the pixel's resonant frequency. The firmware detects the
absorption of photons as a sudden jump in the phase of a pixel's probe
tone, followed by an exponential decay back to the baseline phase
(Figure \ref{fig:photonPulses}).  The phase timestream from each pixel
is passed through a matched filter made with a template phase pulse
specific to the pixel.  The firmware also keeps track of the baseline
of the phase for each pixel with an infinite impulse response (IIR)
low-pass filter.  When a peak is detected above a threshold value, the
firmware generates a photon data packet to be saved to disk. By
design, a $100 \us$ dead time ($10\us$ in the most recent ARCONS
implementation) follows each photon detection, during which no other
photon events will be recorded to disk.  This is to prevent false
photon detections from phase noise during the decay back to the
baseline.  The effect of this dead time on the linearity of the
pixel's response is corrected (See Section
\ref{sec:linearity}). The phase timestream itself is not saved
  while observing due to the prohibitively high storage and data-transfer rate
  requirements.

\subsection{Raw Data Output}
\begin{figure} \epsscale{1.12}
\epsscale{1.12}
\centering
\plotone{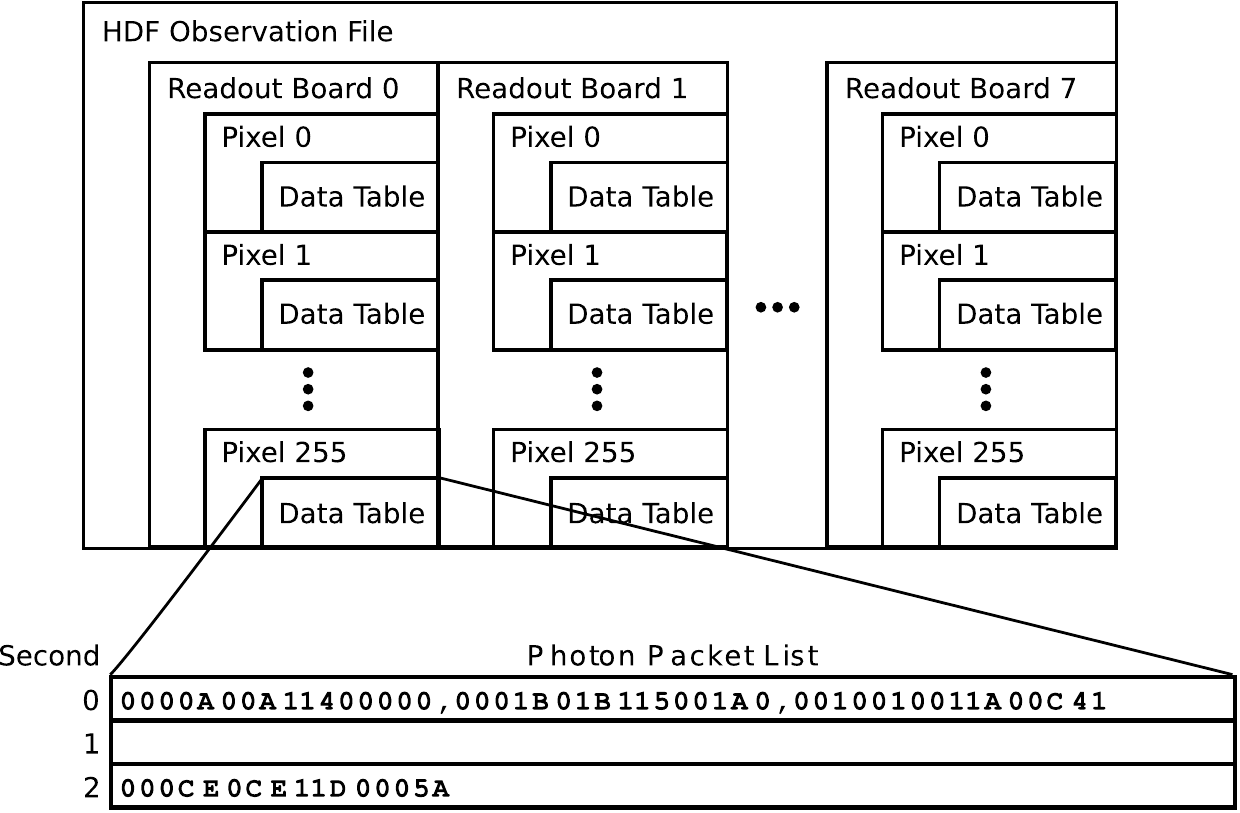}
\caption{Organization of the raw data file created during
  observations. The file is organized first by readout board, then by
  pixel.  Each pixel is allocated a data table, in which each row
  corresponds to a second of observation time. Each row in the data
  table has a photon data packet for every photon detected for the
  pixel during the corresponding time, stored as a ragged array.}
\label{fig:obsFileDiagram}
\end{figure}

\begin{figure} \epsscale{1.12}
\epsscale{1.12}
\centering
\plotone{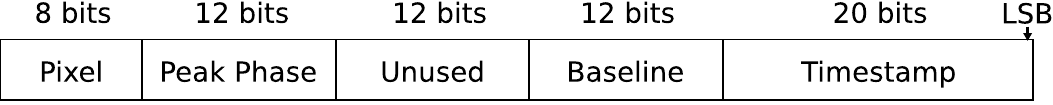}
\caption{Structure of the 64-bit packet generated during a photon detection.
The pixel number is only 8 bits because it represents an index into a list of
pixels assigned to a readout board, each of which can read out 256 pixels. The
least significant bit (LSB) is shown as the rightmost bit.}
\label{fig:photonPacketDiagram}
\end{figure}

The raw photon data packets are stored in an HDF5 file,
organized by readout circuit board and pixel (Figure
\ref{fig:obsFileDiagram}), with a single data table allocated for each
pixel.  Each row of a pixel's data table contains a list of 64-bit
packets representing the photons detected in one second of exposure.
The packet records the pixel where the photon landed, the phase value
at the peak caused by the photon in radians, the concurrent phase
baseline, and the arrival time of the photon relative to the beginning
of the current second (Figure \ref{fig:photonPacketDiagram}).

\subsection{Barycentric Photon Arrival Times}\label{sec:barycenter} To
determine a photon's arrival time, first the start time of the
observation is extracted from the raw HDF5 data file's header.
Initially this is formatted as a Unix timestamp, but it is later
converted to UTC or to a Julian Date.  The 20-bit timestamp in a
64-bit photon packet is the number of output sample time steps since
the beginning of the second, where a ``time step'' is currently
$1\us$.  The second in which the photon arrives relative to the start
time of the observation corresponds to the row number in the pixel
data table where it is stored.  The complete absolute time is found by
adding the start time of the observation, the seconds elapsed since
the beginning of the observation, and the photon timestamp in
$\mu\mathrm{s}$.  The timestamps are then corrected for a $41\us$
delay in the firmware (previously measured by illuminating the
detector with a light emitting diode triggered by the pulse-per-second
signal from our GPS clock reference). 

In addition, there are occasions when a readout board is slow to
offload its data to disk on completion of integration for an
observation file, and may be delayed at this point while the other
boards have begun collecting data for the next file.  In this case,
the slow board will start taking data for the new file exactly one
second later than the others (readout boards always start recording
data at the beginning of a second). The timestamps from this board
require a one-second offset, since the start time for this board is
one second off from the start time read from the header. Log files
written during observation indicate when these second offsets are
needed. There is an additional problem in that the slow board
therefore misses a second of data that the others have recorded. To
compensate, we generally throw out data where we do not have all the
boards reading out simultaneously, effectively adding a second of down
time between observation files. This issue has been greatly improved,
so that in our last observing run, instances of slow boards were very
rare.

After applying corrections, timestamps can be corrected to
solar system barycentric times with the TEMPO2 pulsar timing package
\citep{Tempo2} using a custom plug-in that allows TEMPO2 to treat
photons individually. TEMPO2 takes parameter files indicating the
location of the observation and the ephemeris of the object observed.

\subsection{Beam Mapping}\label{sec:beammapping} During the MKID design
process, each pixel is assigned a unique resonant frequency by systematically
varying the design of the capacitor for each pixel. The pixels are then
assigned to positions on a rectangular grid according to an algorithm that
randomizes their locations, maximizing the physical distance between
pixels with similar resonant frequencies to avoid cross-talk. In theory, this
creates a known mapping of pixel resonance-frequency to chip location. In
practice, however, random non-uniformities in the superconducting
critical temperature across the array shift the resonant frequencies in
unpredictable ways.  We must measure the actual resonant frequencies after
fabrication and then empirically determine which frequencies correspond to
which pixel location.

We first find the resonant frequencies of all the
pixels by using a vector network analyzer to sweep probe signals through the
chip across all frequencies in a few-GHz band, and tag the dips in signal
transmission that signify a resonance. The resulting list of
resonances is used to map readout frequencies to indexed pixel numbers.  To
determine each pixel's location, we systematically illuminate the pixels using
an \textit{X--Y} translation stage holding a photomask patterned with
two long, orthogonal slits, illuminated with a beam of light from the
backside. We scan these strips of light across the array one at a time, then
use the time stream of each resonator's response to precisely determine the \textit{X--Y}
position when it was illuminated. The end result is a beammap file which
assigns a pixel number to its corresponding resonant frequency and physical
grid location. The beammap is loaded as a two-dimensional (2D) look-up table during
data analysis to map each pixel to its proper location in
the image.

The beam-mapped raw data provide the input for the data reduction
pipeline, which eventually produces fully calibrated photon lists. The
steps involved in this calibration are detailed below.

\section{Pipeline components}\label{sec:pipeline}

\subsection{Cosmic-ray Cleaning}\label{sec:cosmicrays}
We use the term ``cosmic ray'' to describe events where photons arrive
in different pixels nearly simultaneously.  We attribute these events
to radioactive decays and muons interacting with the MKIDs' silicon
substrate. When a cosmic ray interacts with a MKID or the crystal
substrate, it deposits much more energy than an optical photon.  This
creates a burst of phonons which are seen by many of the MKIDs in the
array as false high energy photon events. We therefore see a broad ``flash''
across much (or all) of the array, in contrast to the cosmic ray
tracks commonly seen in conventional CCD images. These rare events are
eliminated using the coincident arrival times with a loss of less than
0.1\% of the total exposure time.  The algorithm calculates time
intervals within a data file that are near cosmic ray events and then
excludes data in those time intervals from further analysis.  The
following parameters control the algorithm:
\begin{description}
\item[\tt stride] the number of consecutive output sample time steps
  combined to form one ``time bin'';
\item[\tt threshold] the minimum number of photon packets in a time
  bin to determine whether a cosmic event is detected;
\item[\tt width] the number of time bins to exclude around a
  detected cosmic ray event.
\end{description}

We read in the time for all of the photon packets from all of the
pixels in a data file.  We form time bins from consecutive time
samples, starting at the first time step.  We then form overlapping
bins by starting at time step {\tt stride}/2.  For all of the time
bins that have more than {\tt threshold} photons, we calculate a time
interval to exclude.  The number of apparent photon detections as a
function of time increases quickly in these events, with a longer
decay tail.  We exclude {\tt width} time bins before the cosmic ray
event and twice that length after the end of the time bin.  We form
the union of all of the excluded intervals of a data file and use this list of
intervals to mask cosmic rays.

\begin{figure}[h]
\epsscale{1.12}
%\plotone{cosmic-tune-eps-converted-to.pdf}
\plotone{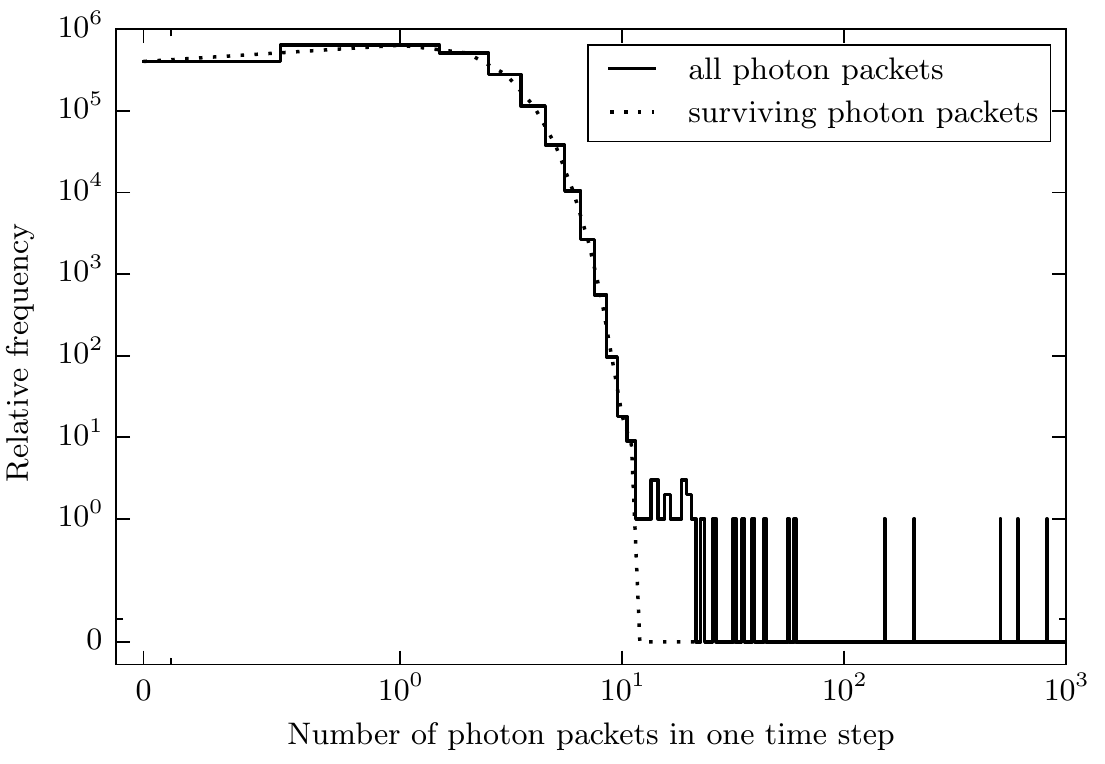}
\caption{Distribution of number of photons per time sample.  The solid
  line is all data (no cosmic ray masking) and the dotted line is
  after masking time steps near cosmic ray events.}
\label{fig:cosmic-tune}
\end{figure}

Figure \ref{fig:cosmic-tune} shows the distribution of the number of
photons per time step for a typical data file.  The solid line is for
all photon packets, with no masking.  Note that the population is
dominated by a Poisson distribution, with $\mu \sim 0.5$. Cosmic ray
events are in the tail of the distribution with population $ > 10$
photons/time step.  We mask using time intervals calculated with {\tt
  stride} = 10 ($10\us$ for our current $1\us$ time step), {\tt
  threshold} = 15 photon events, and {\tt width} = 10 time bins
($100\us$).  This eliminates events in the tail and excludes $0.02 \%$
of time intervals.

\subsection{Bad Pixels and Exposure Time Masking}\label{sec:timemasks}
It is found that some pixels in the array go ``hot'' from time to
time, registering false events and reporting very high count rates,
commonly over periods of order seconds (for the ARCONS 2012 data). We
refer to these as ``hot'' pixels by analogy with the phenomenon of the
same name encountered in conventional CCDs, albeit in the case of CCDs
they tend to be more often permanent than transient features. In our
case it is suspected that the effect is due to an interaction of the
MKID with free electrons in the silicon wafer, which should be
eliminated in future MKID designs.

In order to catch any such temporarily anomalous pixels, we use an
algorithm similar to that employed in the {\tt badpixfind} task of the
\textit{XMM-Newton} X-ray data pipeline.\footnote{\textit{XMM-Newton} Science Analysis
  System, Users Guide to the XMM-Newton Science Analysis System, Issue
  9.0, 2012 (ESA: XMM- Newton SOC). See also
  \url{http://xmm.esac.esa.int/sas/}.} Because the point spread
function (PSF) of the telescope is oversampled by the instrument, any
real astrophysical light source must present a minimum finite size on
the detector. If the ratio of the flux in a pixel to the median flux
in a surrounding box (say, $5\times 5$ pixels) is significantly higher
than could be expected for an astrophysical point source on the
detector, then the apparent PSF is too narrow to represent anything
real, and the pixel is flagged as hot and masked out. The threshold
ratio (after accounting for sky background) is determined as the ratio
of the peak of a model 2D Gaussian function to the median
of the same function in a surrounding box of equivalent size.
The FWHM of the Gaussian is set to
  match the expected seeing, and an additional ceiling is added to the
  threshold which accounts for both photon shot noise and the standard
  deviation of the measured sky background (typically
  at the $3\sigma$ level). Using the standard deviation of the sky background helps
  to account for any internal instrument noise (due to variations in
  QE, etc.). Since neighboring bad pixels will skew the local median
used for flux comparison, this process is iterated until there is no
change in the mask or a maximum number of iterations is reached.

The flagging is performed in regular short integration time-slices
(typically $1\seconds$) for each observation file, in order to capture
the time variability of bad pixels and minimize the amount of good
data that is unnecessarily flagged. The resulting masks are recorded
as a list of bad time-intervals for each pixel, creating a single
time-mask file for each input raw data file.

``Dead'' pixels---those recording no counts at all---are also
flagged. In all cases, the reason for flagging is also associated with
each recorded bad time-interval in the time-mask file. Current
possible ``reasons'' are confined to dead pixels, hot pixels, and,
optionally, ``cold'' pixels which may give abnormally low, but
non-zero, count rates; there is facility to add further flagging as
the pipeline matures.\footnote{In certain cases, mislocated pixels
  resulting from beammap errors (Section \ref{sec:beammapping}) may
  cause a pixel from the core of the PSF of a star to be mismapped
  into a region of empty sky. In such a case, the pixel will also be
  flagged as ``hot''. Usually these mislocated pixels can be found and
  corrected in the beammap.}

The algorithm is able to account somewhat for variability in QE from
pixel to pixel, and the raw QE is reasonably uniform. However, it is
better able to discriminate hot pixels with flatfielded data, where
the noise levels are lower (see Section \ref{sec:flatcal}). Hot-pixel
flagging can therefore be run either before or after the flatfielding
has been performed (or both), depending on requirements.

\begin{figure*}[tb]
\epsscale{1.12}
\centering
%\plotone{badPixels-eps-converted-to.pdf}
\plotone{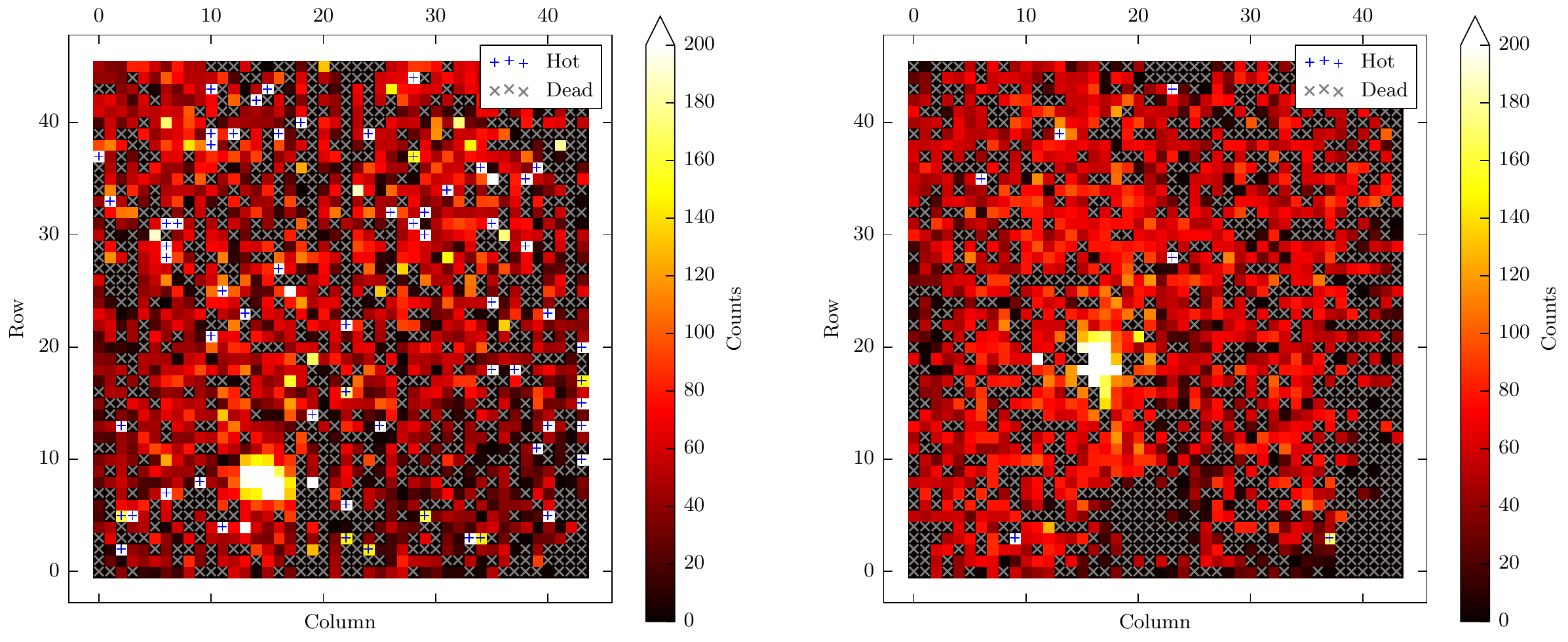}
\caption{Bad pixels detected in example $1\seconds$ time-slices of raw
  data from ARCONS. \textit{Left:} AM CVn binary SDSS J0926+3624,
  taken 2012/12/09 (UT). \textit{Right:} compact binary SDSS
  J0303+0054, taken 2014/10/22 (UT). Pixels with high flux within the
  vicinity of astrophysical sources that are consistent with the PSF are not
  flagged, while others outside source vicinities
  are flagged as ``hot.'' Improvements to the ARCONS array in 2014 led to substantially
  reduced hot-pixel behavior. Additional dead pixels in the 2014 data
  are are caused by magnetic flux trapping in the superconductors
  during array cool-down (the ``hole'' at bottom center), and by a
  fault in one of the eight readout circuit boards (bottom right of
  the array). Both these issues are correctable in future arrays.}
\label{fig:badPixels}
\end{figure*}

The masks for two $1\seconds$ time slices of typical raw data files
from 2012 and 2014 are shown in Figure \ref{fig:badPixels}, showing
pixels that have been flagged as either hot or dead. For the same
observation file from which the 2012 mask was derived, $\approx 8\%$
of the pixels were flagged as exhibiting hot-pixel behavior at some
point during the $300\seconds$ duration of the data file (of which
$\approx 0.2\%$ were flagged as hot for the entire $300\seconds$). The
distribution of hot intervals began with a peak at $1\seconds$
(the sampling integration time) and tailed off toward longer
durations. $31\%$ of the pixels registered as ``dead'' for the
duration of the exposure (including un-mapped pixels,
etc.). Approximately $4\%$ showed temporary dead behavior, which seem
largely to be the result of Poisson statistics for pixels which are
registering extremely low count rates, either because they are faulty
or because of exceptionally low quantum efficiency.

The new array used in the ARCONS 2014 run showed a significant
improvement in hot pixel behavior. In the 2014 example image shown,
only $\approx 0.5\%$ of the pixels showed hot behavior at some point
in the observation file (0.15\% for the entire duration). 41\% were
dead, an increase over the previous observations. This was due to
magnetic flux trapping in the superconductor during cool-down and to a
faulty readout circuit board, both of which can be corrected in future
observing runs. Most of the remaining ``dead'' pixels in the arrays
from both years result from collisions in the resonator frequency
domain.  This occurs because non-uniformities in the TiN film of the
MKIDs shift the resonant frequency of pixels away from the design
resonant frequency.  When two or more resonators are less than
500\,kHz away from each other in resonant frequency we can only read
out one of these resonators, leaving the others not addressed by the
readout.  More uniform superconducting films for MKIDs are currently
being intensively investigated, and should drastically decrease the
number of dead pixels.

The masking process does not capture {\em all} hot pixels:
some may be hot but have flux levels which are marginally within the
noise in a given time slice, and thus not identified. In principle,
hot pixels have other characteristics which may be used in the future
to further refine the flagging: the spurious counts often---though
not always---have smaller event pulse heights, which the pipeline
interprets as longer wavelength, so that spectral characteristics may
help identify the behavior. Similarly, hot pixels often tend to switch
``on'' or ``off'' very suddenly (or change between various flux levels
in a step-wise manner), which would be unusual for a true
astrophysical source. Conceivably this timing and spectral information
can be used to better identify bad pixels, particularly if it can be
determined that neighboring pixels are not exhibiting the same
behavior (since this would again tend to rule out an astrophysical
source). It is also hoped that array development will mitigate many of
these effects.

Future improvements to the detection of permanent bad pixels may be
obtained by creating additional masks in a way similar to traditional
bad pixel masking for CCDs. By dividing two different flatfield
calibration images at different flux levels (e.g., sky flats taken at
different times during twilight), it is possible to identify outlying
pixels whose flux response does not behave in proportion to the
majority of well-behaved pixels.

\subsection{Wavelength Calibration}\label{sec:wavecal}
\begin{figure}
\epsscale{1.12}
\plotone{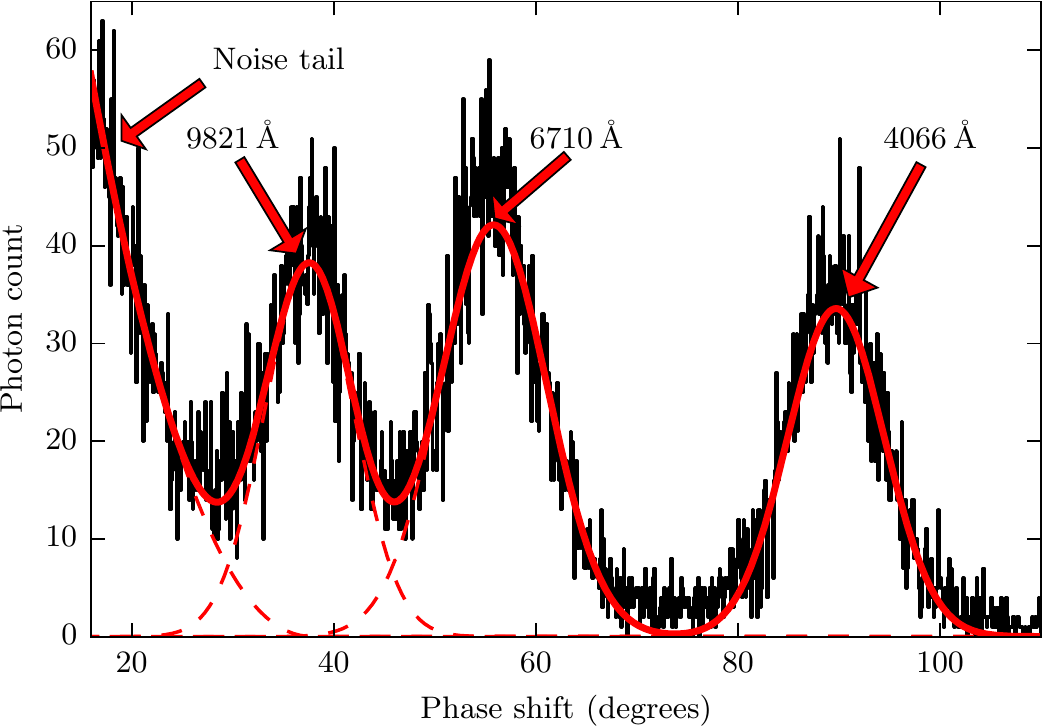}
\caption {Histogram of a pixel's resonator phase shifts due to photon events
  from simultaneous exposure to a blue, red, and IR laser. Gaussians
  are fit to each laser's peak, and a parabolic fit to the locations
  of those peaks is used to convert phase shift to photon energy. A
  noise tail arises possibly from some combination of resonator
  baseline phase noise, an IR leak in the cold-filter transmission
  function, and photon interactions with the silicon substrate.}
\label{fig:wvlcal_phaseHist}
\end{figure}

Wavelength calibration involves determining the mapping from the
photon-induced phase shift in a pixel's microwave probe tone to the
photon's corresponding wavelength. For the purposes of calibrating the
wavelength response for each pixel, $300\seconds$ wavelength
calibration exposures are taken regularly throughout the night. These
are created by uniformly illuminating the array with the combined
light from a blue, a red, and an IR laser (currently at wavelengths
4066, 6710, and $9821\ang $ respectively). For each pixel, a histogram
of the phase shifts for every photon event in a given calibration
exposure is accumulated, which shows three peaks corresponding to the
three laser wavelengths (Figure \ref{fig:wvlcal_phaseHist}). Gaussian
functions are fitted to each of these peaks to find the locations of
their respective maxima, which are taken to be the best estimate of
the phase shift for the corresponding laser wavelength. A parabolic
fit to these three phase values as a function of wavelength
then gives a continuous estimate of the function mapping phase shift
to wavelength for that pixel. The width of the Gaussian fits gives an
estimate of the effective resolving power at each of the three
wavelengths. A number of cuts and checks are performed to ensure a
sensible fit, and any pixel which does not pass the checks is marked
as failed. The results of the fits are stored as a calibration
  file, containing, for each pixel, the parabola
coefficients, the wavelength range of validity of the calibration, any
flags associated with the calibration (e.g., for pixels where the
calibration fails), and a record of which ROACH readout board is
associated with the pixel. A long wavelength cutoff is
  established at the point the count rate from the IR laser is equal
  to the noise, allowing noise counts to be ignored in later
  processing. Resolutions are typically $R\sim6.5$ at
$4000\ang$, but range from $R=3$ to $R=10$ for different
pixels.

Properly accounting for photon pileup effects in the wavelength
calibration data was found to be essential for a good solution.
Failure to do so leads to significant broadening and skewing of the
three histogram peaks, leading to systematic wavelength calibration
errors and an apparent reduction in spectral resolution. This results
from systematic errors in the measured phase baseline and phase shift (see
Section \ref{subsec:reddening}). Such effects are corrected by
removing photons that occur within a nominal $1\,\mathrm{ms}$ of the
previous photon. The timescale of $1\,\mathrm{ms}$ can be shortened
but it is more than long enough to remove any photon pileup effects.

\begin{figure}
\epsscale{1.12}
\plotone{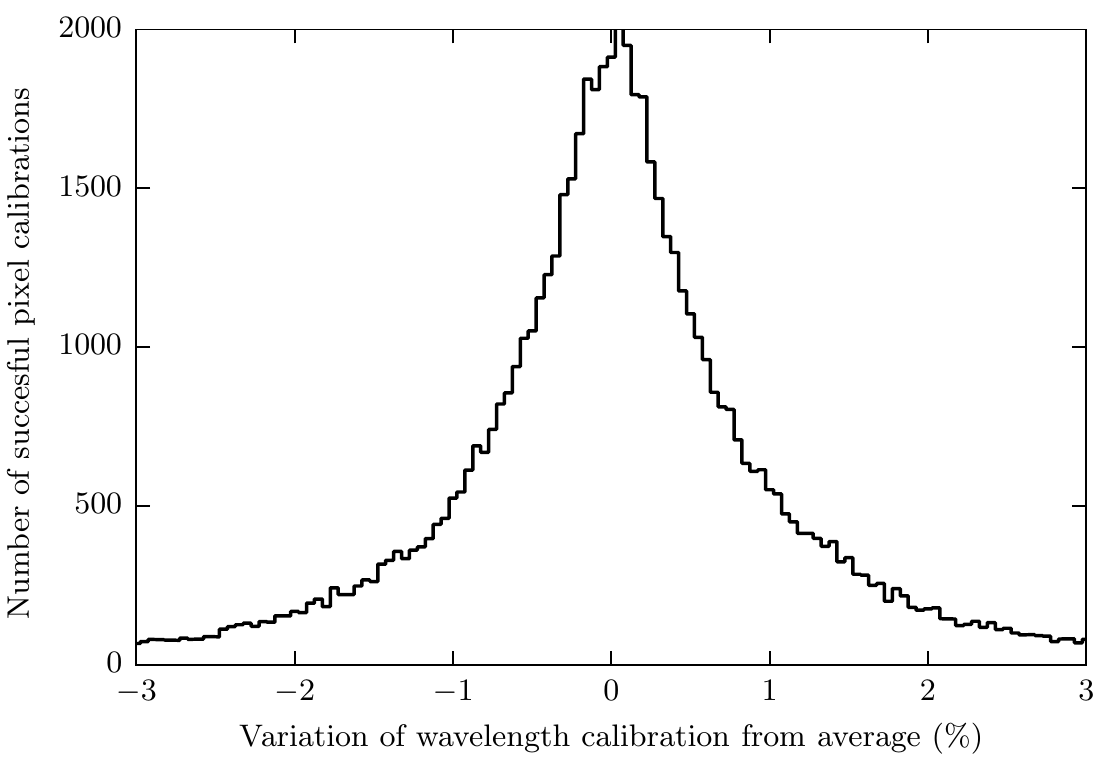}
\caption {For each wavelength calibration exposure taken during a
  night, we find slightly different phase shifts caused by blue
  ($4066\ang $) photons (See Figure \ref{fig:wvlcal_phaseHist}). We
  determine the averaged master phase shift for each pixel and then
  calculate the variation for each single phase shift against the
  average. Here, we histogram this variation for every pixel and every
  wavelength calibration exposure to probe the overall array's
  wavelength calibration stability during the night. The distribution
  is centered around zero, indicating no apparent array
  systematics. The FWHM is about 0.5\% which is small compared to the
  energy resolution of an individual photon of $\Delta E/E = 15\%$. }
\label{fig:wvlcal_fluct}
\end{figure}

Pixels with a valid wavelength solution have their gaussian fit
parameters saved in a calibration ``drift'' file. The
stability of the wavelength calibration is determined from the pixel's
response to the blue laser light over time. We find that throughout
the night, most pixels' peak phase shifts for blue photons (as
determined by the Gaussian fitting procedure) vary by less
than 0.5\% (Figure \ref{fig:wvlcal_fluct}). If the hardware creating
the pixel probe tones is restarted, peak phase responses can drift
systematically by as much as 4\%. A detector array temperature change
of about $5\mK$ will systematically shift the peak phase response for all
pixels by about 5\%. Since the temperature is controlled to better
than $0.05\mK$, temperature drift is not expected to be a significant
source of error.

Since the pixel's wavelength calibration is quite stable over a night
(0.5\% compared to $1/R=15\%$), we average several wavelength
calibrations into a single ``master'' wavelength calibration file.
The UT times for which the master wavelength calibration file is valid
are also saved. The wavelength of an observed photon is then
determined by looking up the fit coefficients for the photon's pixel
in the corresponding master wavelength calibration file, and using
these to map phase shift to wavelength. The use of a master wavelength
calibration has two main advantages. First, by averaging solutions
from several wavelength calibrations we can get a more accurate
calibration, assuming there are no significant systematic shifts in a
pixel's phase response. Second, if a pixel has a bad wavelength
solution for a specific wavelength calibration we can recover a
solution by averaging the remaining successful wavelength
calibrations.

\subsection{Flatfield Calibration}\label{sec:flatcal}
The variations in response from one pixel to another are normalized
using sky exposures during twilight before sunrise (we do not observe
evening twilights to avoid accidentally heating up the detector array
with too much light at the beginning of the night).  Over the course
of morning twilight, the sky's brightness and spectrum change
dramatically, so the twilight data are broken into exposures of
duration $20\seconds$ each, and data are only used until the average
count rate rises above 800 counts per second per pixel.  The
brightness and spectrum of the sky are relatively stable during each
of these exposures.  Each of the $20\seconds$ data chunks is passed
through the hot pixel module.  Then, for each of the data chunks and
each of the pixels with valid wavelength calibrations, the detected
photons are binned into wavelength bins with widths equivalent to
$0.1\ev$.  For each wavelength bin, the median count rate over all
pixels is taken.  This median is divided by the count rate in each
wavelength bin in each pixel, yielding a set of weights, indexed by
pixel and wavelength bin, for each of the $20\seconds$ data chunks.
The weights of the different exposures are combined in a (weighted)
15\%-trimmed mean, to yield a final wavelength-dependent flatfield
weighting for each pixel.

To apply the flatfield calibration to an observation, the observed number of
counts for each pixel and wavelength bin is multiplied by its respective
weight.  This normalizes data from multiple pixels with respect to the average
response, so that comparisons across pixels can be made validly.  This effects
the traditional CCD ``flatfielding,'' but at every wavelength.  When applied to
observations of the sky, the most obvious effect of the flatfield calibration
is to remove a gradient in pixel response thought to be caused by optical
effects.  This gradient is found in data from our 2012 observations but not in
our 2014 observations, done with a different detector array and re-aligned
optics.

To characterize how well the flatfield calibration flattens the pixel response,
we separate the large-scale effects of uneven illumination and intrinsic
pixel-to-pixel variations by creating an alternative set of weights that
corrects uneven illumination, but not pixel-to-pixel differences.  To do this, we
first smooth each twilight wavelength slice image with a sigma-clipped mean,
and then proceed, as in our normal flatfielding, to take the median of the
image and divide it by the image to generate the needed weights.  The smoothed
version of the wavelength slice image captures the large scale spatial
variations in illumination but does not capture any pixel-to-pixel variation.

\begin{figure}
\epsscale{1.12}
\plotone{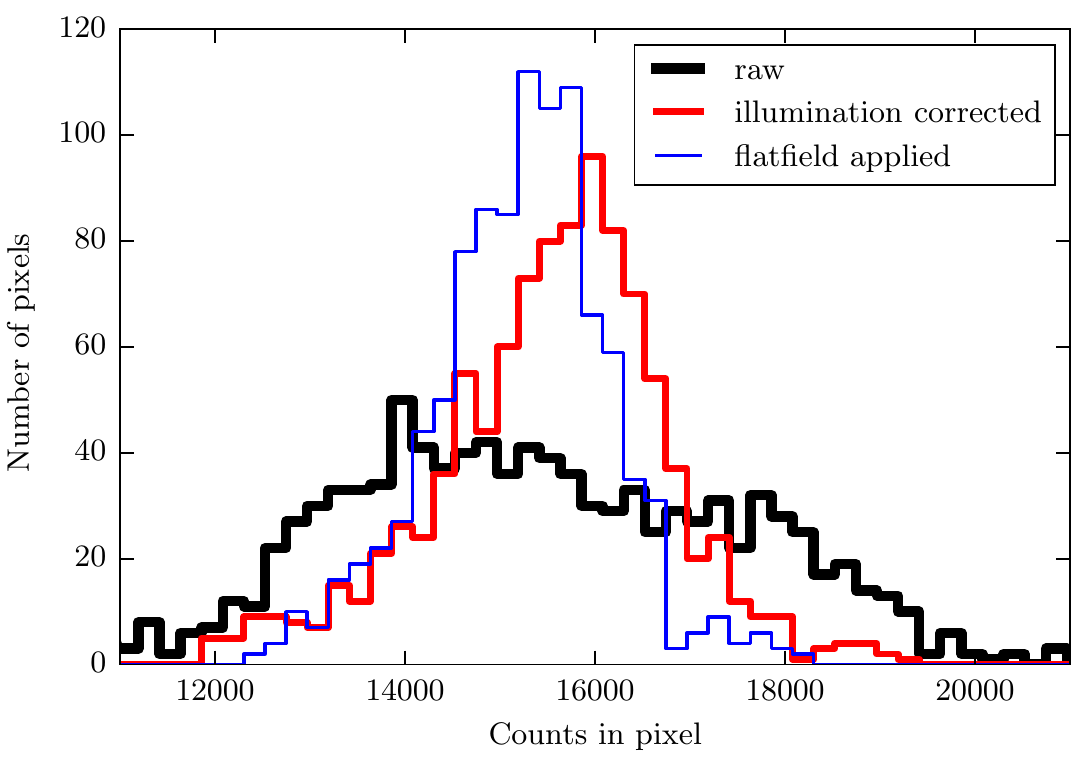}
\caption{Distributions of pixel counts for images made from a
  five-minute sky exposure with no pixel response correction,
  illumination correction only, and with flatfield calibration
  applied.}
\label{fig:flatHists}
\end{figure}

We can then characterize the two sets of weights by comparing the
results of applying either of them to a five-minute night-sky
exposure.  We make images from the exposure showing the counts in the
wavelength range 4000--$11000\ang$ in each pixel
without any weights applied, with the normal flatfield weights
applied, and with the illumination weights applied. The distribution
of counts for these images is shown in Figure \ref{fig:flatHists}.
The distribution for the raw image does not have a sharp peak, because
the large scale gradient smears it out.  We compare the
flatfield-calibrated and illumination-corrected distributions by
taking the FWHM and converting it to a
standard deviation, $\sigma$, with the assumption that the
distributions are roughly gaussian.  We find that the
illumination-corrected and fully flatfield-calibrated distributions
have $\sigma$ values of 6.1\% and 5.0\% of the median count rate,
respectively.  These values are relatively close given the uncertainty
involved in computing them.  For the median count rate in a pixel, we
expect photon shot noise to contribute to the width with
$\sigma=0.8\%$.  If the flatfield calibration were to remove all
pixel-to-pixel variation, we would expect the count distribution to
have this standard deviation.  Instead the flatfield calibrated
distribution has a standard deviation that is comparable to the
illumination corrected distribution.  So, the flatfield calibration
currently accounts for illumination correction, but does not
significantly improve pixel-to-pixel uniformity.

\begin{figure}
\plotone{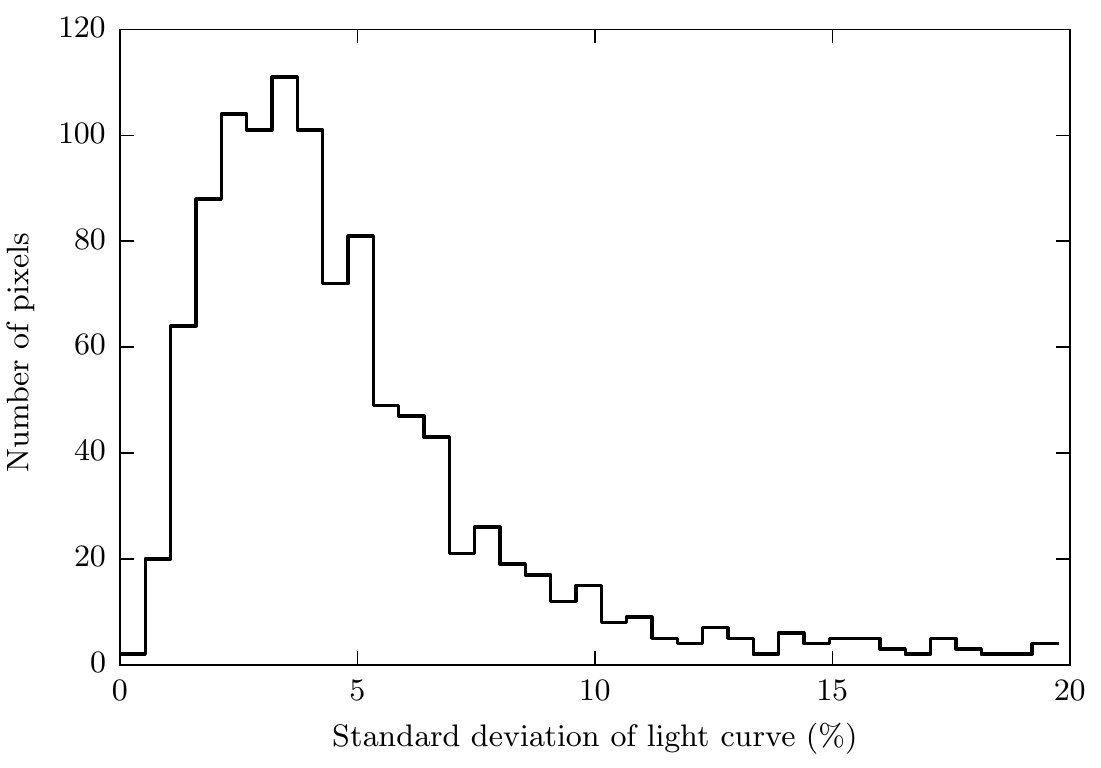}
\caption{Distributions of standard deviations of pixel light curves of
  $2.5\hr$ of night-sky exposures. The standard deviation expected
  from photon shot noise would be less than 1\% for all pixels.}
\label{fig:pixelStabilityHist}
\end{figure}

To explain this, we checked the stability of a pixel's count rate with
time.  For this test, we used $2.5\hr$ of night-sky exposures,
divided into segments of 150\,s each. A light-curve is made for each
pixel, which is normalized by a reference light-curve made of the
median counts over all pixels for each time segment.  The standard
deviation of the normalized light curve is then taken for each pixel.
The distribution of standard deviations for all pixels is shown in
Figure \ref{fig:pixelStabilityHist}.  For all pixels, the standard
deviation expected from photon shot noise is less than 1.0\%.  Nearly
all pixels have standard deviations above this, indicating that there
is some instability in the flatfield with time. The peak is close
to the standard deviation of the flatfield-calibrated count
distribution from Figure \ref{fig:flatHists}, which accounts for the
limited ability of the flatfield calibration to completely flatten
pixel response.

\begin{figure}
\plotone{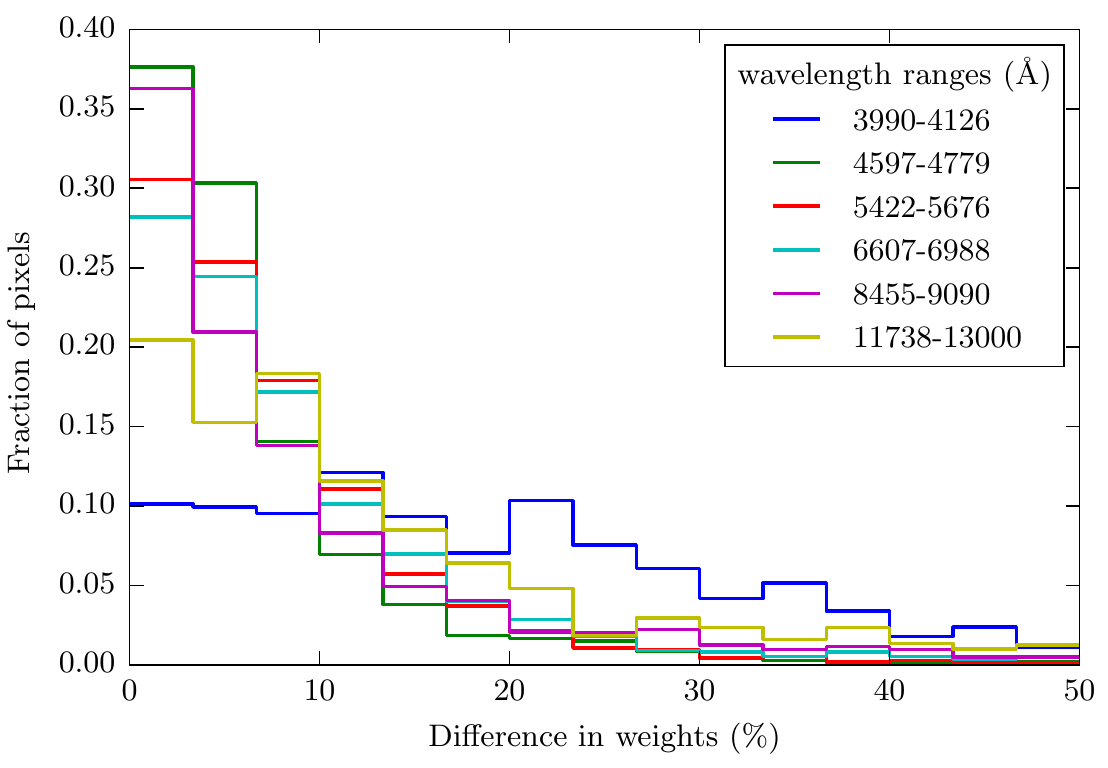}
\caption{Distributions of percent differences between flatfield weights from one
    night's twilight and the next for a selection of wavelength bins.}
\label{fig:flatVariations}
\end{figure}

We also investigated the stability of a pixel's calibrated flatfield weight
from one night to the next.  Figure \ref{fig:flatVariations} shows the
distributions of percent differences between flatfield weights calculated for
one night's twilight and for the next night. There is a sizeable fraction of
pixels whose calculated weights change by 10\% or more, particularly in the
bins with the shortest and longest wavelengths.  This is possibly due to changes
in those pixels' wavelength calibrations from one night to another.

In a few pixels, the flux observed during twilight is not consistent with the
QE observed indirectly at other times of the night, mostly likely due to those
pixels having unreliable wavelength calibrations.  These pixels receive
flatfield weights either much lower or much higher than they should be
assigned.  The effect is that after the flatfield calibration is applied, these
pixels appear hotter or colder than they should be.  Because of this, and in
light of the fact that in its current state the flatfield calibration does not
generally improve pixel-to-pixel uniformity, we currently only apply the
illumination correction.  The weights calculated for this correction do not
have extreme changes from one pixel to a neighboring pixel, and in general, these
weights also change less from one night to the next than weights calculated
without smoothing.

\subsection{Spectral Response Calibration}\label{sec:speccal}
Once the pixel responses are normalized relative to one another by the
flatfield calibration, absolute spectral shape is calibrated by dividing an
ARCONS measured spectrum of a spectrophotometric standard star by the
known spectrum of said target.

To extract the raw ARCONS spectrum we start with the standard
processing, masking hot pixels and applying the linearity correction
discussed in Section \ref{sec:linearity}. Then the observation is
broken into an image cube with wavelength dependent flat-field applied
and spectral bands spaced $0.1\ev$ apart.  This band size matches the flat 
calibration binning, and over-samples the median MKID energy resolution by 
roughly a factor of 4. We then PSF fit the stellar image in each spectral band 
and integrate the 2-D Gaussian fit to determine the total flux from the object 
in that band.

\begin{figure}
\epsscale{1.12}
\centering
\plotone{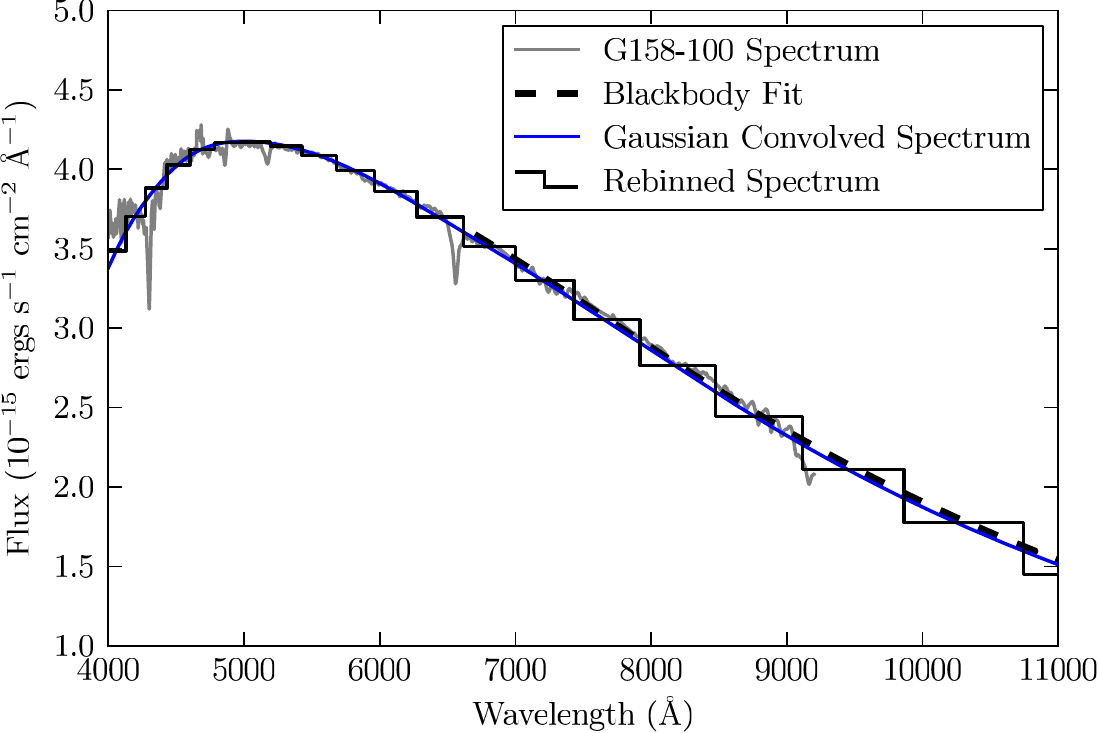}
\caption{To prepare the known spectrum of a spectrophotometric
  standard (solid gray), it is first extended to fill our wavelength
  coverage using a blackbody fit (dashed black), then convolved
  with a Gaussian kernel with width determined by ARCONS' median 
  energy resolution, corresponding to a width of 0.4 eV (blue line). This
  low-resolution spectrum is then resampled by integrating over
  0.1 eV sub-bands, matching ARCONS flat calibration binning, and dividing 
  by bin width in Angstroms to maintain proper $F_{\lambda}$ units (black steps). 
  G158-100 archival spectrum from \url{www.eso.org}.}
\label{fig:FluxCal_Std}
\end{figure}

To prepare the known standard spectrum for comparison we first extend
the near-IR coverage using a blackbody fit, since ARCONS' broad
bandwidth extends to longer wavelengths than most commonly available
spectrophotometric standard data. The pipeline can also accommodate
model spectra covering any desired wavelength range, obviating this step.
We then convolve the high resolution spectrum with a Gaussian kernel of 
width determined by the median energy resolution of the MKIDs (0.4 eV width
for 2012 data) to simulate how the spectrum would appear at ARCONS'
median energy resolution, if QE was 100\% at all wavelengths. This 
low-resolution spectrum is then resampled by integrating 
over sub-bands matching the flat calibration binning and dividing 
by bin width in Angstroms to maintain proper $F_{\lambda}$ units. 
Figure \ref{fig:FluxCal_Std} demonstrates this process on G158-100, a 
dG-K type standard with $V=14.89$.

\begin{figure}
\epsscale{1.12}
\centering
\plotone{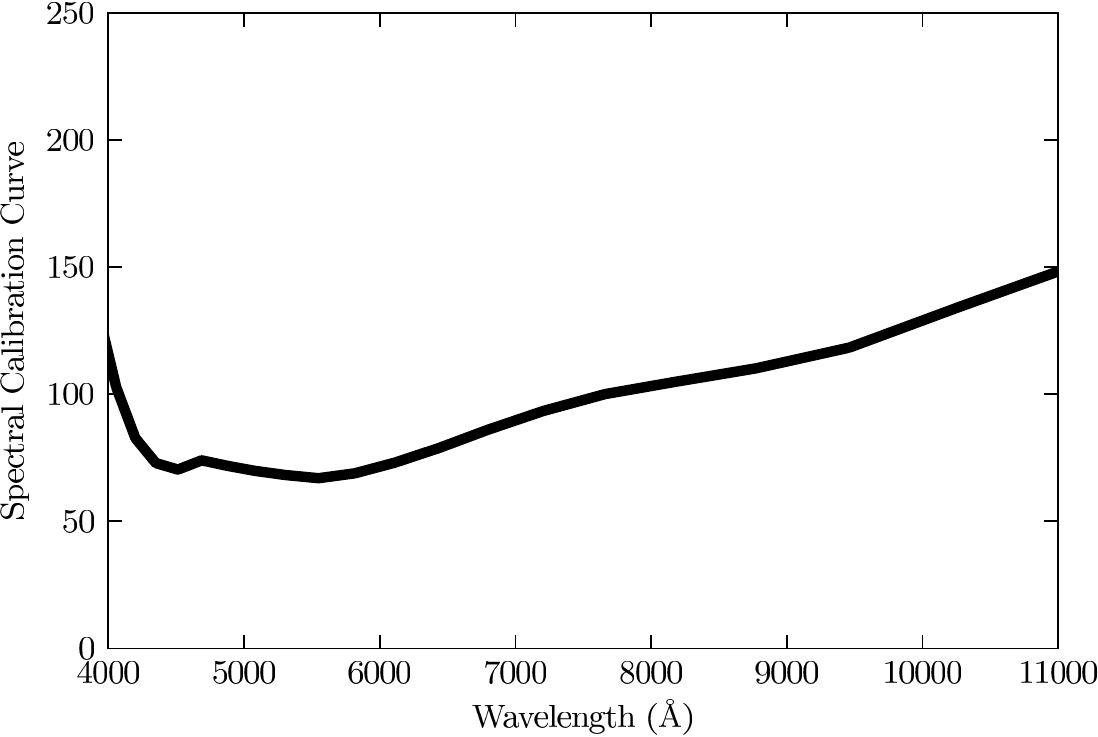}
\caption{Spectral calibration curve from our 2012 run, generated by
  dividing the known spectrum of a standard star by the ARCONS
  measured spectrum (the inverse of the full system throughput). Flux
  calibrated spectra are generated by multiplying flat-calibrated
  spectra by this correction.}
\label{fig:FluxCal_Sensitivity}
\end{figure}

Dividing the two spectra yields a single 
whole-array weighting function designed to flatten the 
spectral response of the detector. When combined with the
per-pixel relative flatfielding weights this spectral calibration also gives 
the overall QE of each pixel as a function of wavelength. Our 2012
spectral calibration curve is shown in Figure \ref{fig:FluxCal_Sensitivity}.

\begin{figure}
\epsscale{1.12}
\centering
\plotone{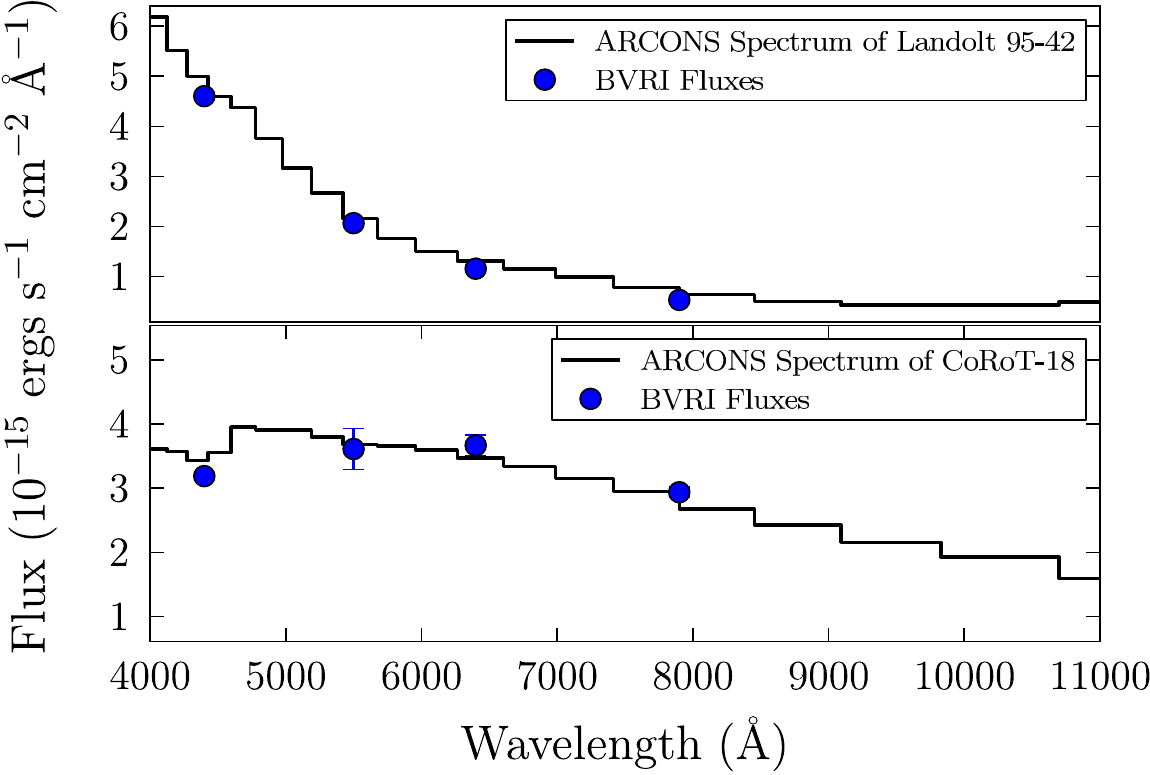}
\caption{Flatfield- and spectral-response-calibrated spectra of known targets,
  Landolt 95-42 \citep{landolt1992} and CoRoT-18 \citep{corot2011},
  demonstrating the accuracy of the spectral shape correction. It is worth noting
  that the cited CoRoT-18 B-band magnitude has no error bars provided.}
\label{fig:FluxCal_Spectra}
\end{figure}

Applied together, the flat weights and spectral response weights yield an 
absolute spectrum for each pixel.
Figure \ref{fig:FluxCal_Spectra} demonstrates the generation of calibrated
spectra on two well-characterized targets, Landolt 95-42 and CoRoT-18, with
known photometry overplotted for comparison \citep{landolt1992, 
corot2011}. For both targets, flatfielded data cubes were generated following 
the same procedure as the calibration star. The spectral
response calibration was then applied as a single whole-array weighting per
wavelength bin, and PSF fitting was performed on these fully-calibrated 
spectral frames. This method of spectrum generation is suitable for
observations with short integrations, such as the ones taken on these targets,
where missing pixels are not filled in by dithering. The general approach
for spectrum extraction from fully processed photon lists is described in
Section \ref{sec:spectroscopy}.

\begin{figure}
\epsscale{1.12}
\centering
\plotone{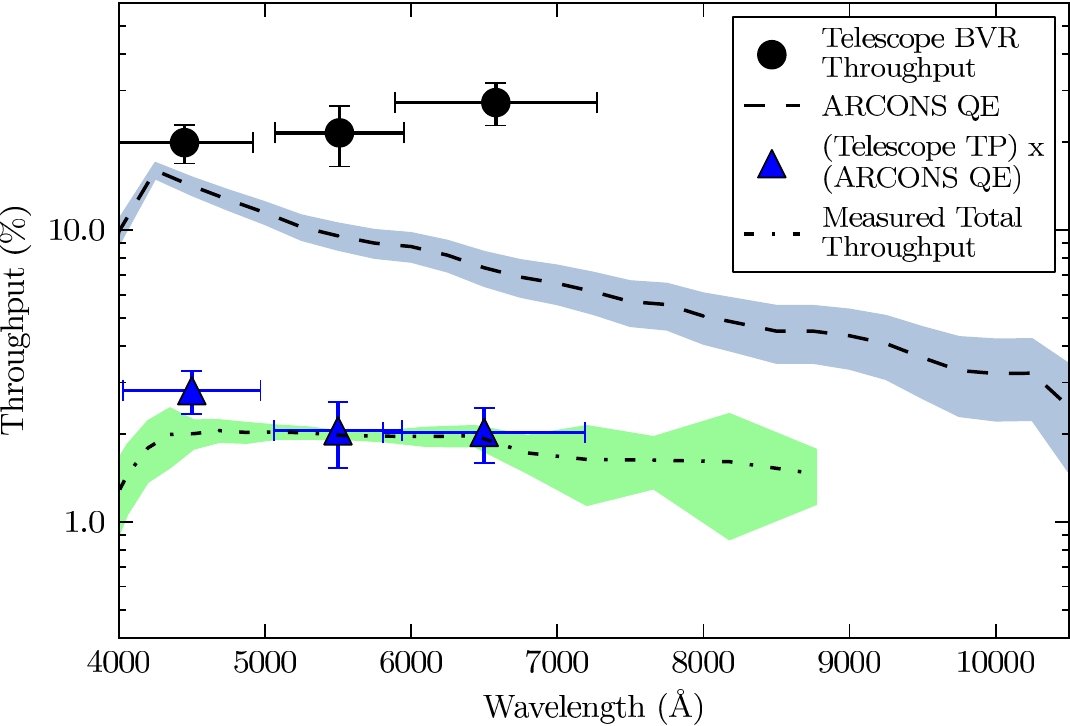}
\caption{A full accounting of light loss performed with data from our
  2013 run. Independent measurements of the Palomar 200" Coude B, V,
  and R throughput (black points) are shown with max/min error bars
  from a few data points. ARCONS lab QE (MKID +
    camera optics, dashed line) is shown with estimated +/-1\% error due
  to potential alignment error of the QE testbed. Telescope throughput
  multiplied by ARCONS QE gives a calculated total throughput estimate
  (blue triangles). This estimate shows good agreement with
  the measured on-sky throughput (dashed-dots). \citep[For more detail on
  MKID and camera-optics QE, see][]{Mazin2012,ARCONS}.}
\label{fig:FluxCal_Accounting}
\end{figure}

As mentioned above, the spectral response calibration also serves as
an integrated throughput measurement including the atmosphere,
telescope, camera optics, and detector. During our 2012 run, we found
the on-sky throughput, as measured during a 5-minute observation of
the spectrophotometric standard star G158-100, to be roughly a factor
of 2 lower than expected \citep{ARCONS}.  When we returned in 2013 we
performed a more careful accounting of our light loss as shown in
Figure \ref{fig:FluxCal_Accounting}. We measured the telescope
throughput independently through Johnson B, V, and R filters using a
UDT Instruments Model 221 Silicon Sensor placed just before the ARCONS
entrance window, read out with a UDT Model S480 flexOptometer.  The
measured throughput values of ~20-30\% are roughly 2 times lower than
we expected, matching previous studies of aluminum coating degradation
over many years \citep{Magrath1997} and accounting for our extra light
loss. We have attributed most of this degradation to the Coud\'e flat
mirror, which has since been re-aluminized and should now be
restored to a pristine state. Due to poor atmospheric conditions on
the 2013 run, and the only available spectrophotometric standard star,
HZ21, being too faint at red wavelengths, we were not able to achieve
an on-sky throughput measurement across our entire bandwidth even with
a factor of 4 times longer integration than in 2012. However,
the data shown here still demonstrate that the combination of
independent telescope throughput and ARCONS QE (optics + detector)
measurements matches the final on-sky throughput. Together, these data
give a complete picture of where light is lost during our
observations.

The QE of the MKIDs is simply the fraction of photons that are
absorbed in the TiN metal.  Since MKIDs are fundamentally sensitive to
energy and do not need to carefully keep track of a photoelectron like
a CCD, it should be possible to use absorbers to increase the QE
significantly.  We are currently investigating vertically aligned
carbon nanotube absorbers \citep[VACNT;][]{Yang2008} that promise nearly
100\% QE across the entire UVOIR wavelength range.

\subsection{High Count-Rate Correction}\label{sec:linearity}
Since MKIDs are single-photon counting devices they can encounter
problems at high count-rates including photon pile-up, data transfer
rates, and pulse triggering. With ARCONS' 2024 pixel array and $1\us$
sample rate, it must apply photon pulse triggering in the firmware for
manageable data storage after a full night of observations. Individual
pixels are limited to count rates of $2500\,\mathrm{cps}$ to avoid any
data transfer backlogs. We find pixels will saturate on
$m_{V}=14$ stars with a throughput like
Figure \ref{fig:FluxCal_Accounting} and nominal $1\farcs 5$
seeing. Brighter stars are observed by defocusing the telescope or
using a neutral density filter. Our firmware is made to trigger on
single photon events and can be nondeterministic under conditions of
photon pile-up. We employ a dead time to avoid triggering on a single
photon twice, and to cut out photons riding on the tail of the
previous photon's phase pulse. This simple event detection algorithm
and subsequent dead time causes a nonlinearity in the detector's
intensity response, and can also systematically lower the apparent
height of a photon's phase pulse.

\subsubsection{Detector Nonlinearity Correction}

\begin{figure}
\epsscale{1.12}
\plotone{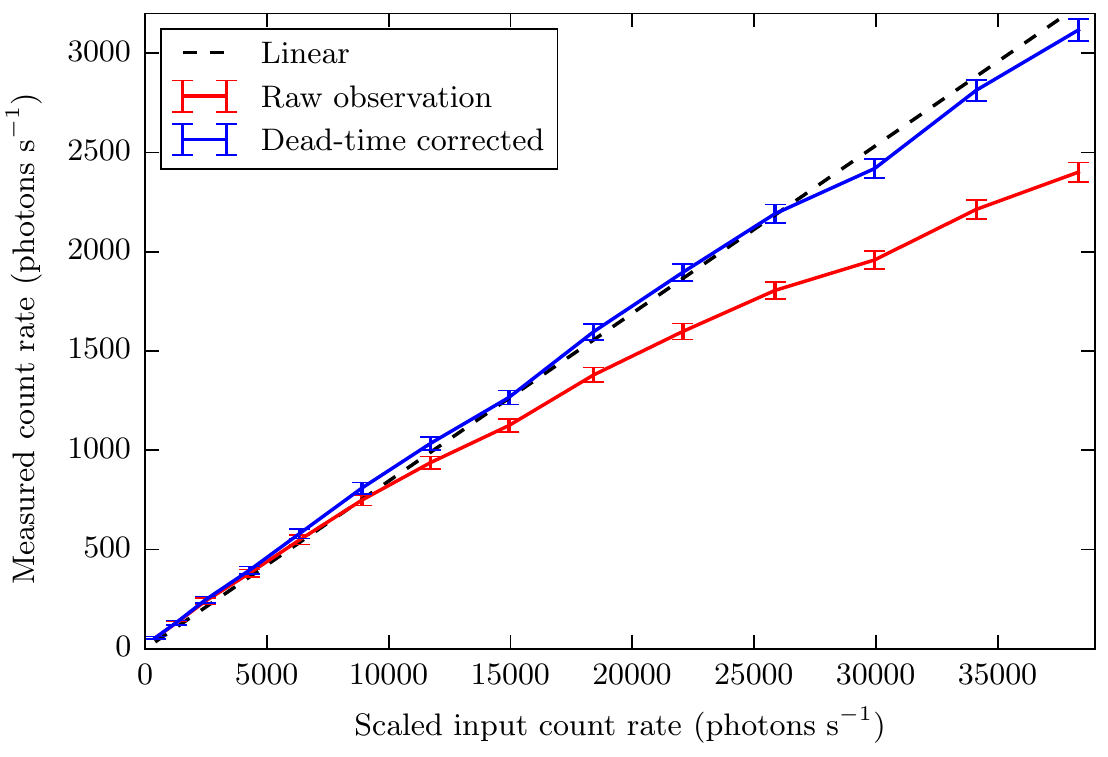}
\caption {ARCONS flux response linearity. Scaled input
  count rate is the flux per pixel incident on ARCONS from a
  monochromator at $5500\ang$, as measured by a calibrated photodiode.
  Measured count rate is the average number of photons recorded by
  ARCONS per pixel. The red connected points show raw data while the
  blue is dead-time corrected. The dashed black line shows the
  expected count rate for a perfectly linear detector with 8.5\%
  quantum efficiency. This figure is reproduced with data from
  \citet{ARCONS}.}
\label{fig:linearityCorrection}
\end{figure}

Photons arriving within a given time (usually 10-$100\us$) of a
preceding photon on the same pixel are locked out by design in the
firmware, creating a non-paralyzing\footnote{Photons arriving during
  that interval do not reset the lock-out; if they were to do so, then
  very high count rates would prevent \emph{any} photon triggers, and
  the detector would be ``paralyzed.'' A non-paralyzing dead time
  instead continues to allow event triggers, but at a rate limited by
  the dead time.} dead time (see Section \ref{sec:pulsedetection}). At
high count rates, this systematically lowers the effective quantum
efficiency. A chunk of time, either within an exposure time or
spanning multiple exposures, can be corrected by weighting count rates
by a factor of $(1-N \cdot \tau / T )^{-1}$ where $N$ is the number of
detected photons in exposure time $T$ for a pixel with dead time
$\tau$. Figure \ref{fig:linearityCorrection} shows the linearity
response before and after dead time correction. Both the flatfield
calibration and spectral response calibration employ a $100\us$ dead
time correction ($10\us$ for more recent 2014 data).

\subsubsection{Wavelength Bias Correction}\label{subsec:reddening}

The baseline of the phase signal from a pixel, recorded with a
low-pass filter in the firmware (see Section
\ref{sec:pulsedetection}), is systematically changed by the phase
shift of an arriving photon. This has the effect of raising the phase
trigger threshold, and hence the minimum trigger photon energy, for a
photon pulse arriving within several hundred microseconds
afterwards. This timescale is set by the timescale of the low-pass
filter. The baseline shift is small enough, and the trigger threshold
is set low enough, that we do not lock out photons within our
wavelength band. However, photons arriving within several hundred
microseconds of a previous photon event appear to yield smaller phase
shifts due to an artificially raised phase baseline, and are thus
systematically biased redward after wavelength calibration. A longer
low-pass filter timescale can make the effects smaller but the
timescale must be short enough to follow real shifts in the baseline
which occur on millisecond timescales. For most objects, with count
rates less than 1000 cps, this effect can be ignored. For high count
rates, the baseline is biased further with increasing count rates, as
increasingly coinciding photon events skew the average phase
signal. Such effects can be corrected by using only photons which have
no other photon events occurring within a nominal period of
$1\,\mathrm{ms}$ prior to their detection (provided the count rate is
not so high that such photons never occur).

\subsection{Astrometry}\label{sec:astrometry}
Flexure and hose-routing constraints imposed by the ARCONS cryostat mean
that ARCONS must be located at the telescope Coud\'{e} focus at
Palomar, and so the field of view (FOV) rotates continuously with
time. Guiding is generally performed in such a way as to keep the
target object in a fixed position on the array, and the rest of the
field rotates around this position in step with the hour angle of the
observation. In order to assign RA and Dec coordinates for each photon
detection so that long integrations can be properly stacked, it is
necessary to establish an astrometric solution. This can be approached
in two ways, using either the main ARCONS detector images, or using
images from the guide camera, depending on the brightness of sources
available in the ARCONS FOV.

\subsubsection{Science Camera Astrometry}

Where a bright source is present that can easily be detected in short
integration times, a pointing origin is first established with a
precise determination of the object centroid pixel location on the
ARCONS array. To find the centroid position, we use the Python package
PyGuide.\footnote{\url{http://www.astro.washington.edu/users/rowen/PyGuide/Manual.html}}
PyGuide works by finding the point on a two-dimensional image that
minimizes the radial asymmetry of the flux, while accounting for
missing data (primarily caused in ARCONS by unassigned and badly
behaved pixels). The centroid location, in pixels, of the target
object is associated with the object's known catalog RA and Dec. The
centroiding precision is $\lesssim 1$ pixel ($\lesssim 0\farcs 5$)
for a $g^\prime=19$ magnitude object at $1\farcs 5$ seeing.

The orientation of the FOV comes directly from the object's hour angle
at the time of observation (with a possible constant offset depending
on the exact physical orientation of the detector at installation).
The FOV orientation, the origin obtained from the
  centroiding, and the detector plate scale ($\approx 0\farcs 44$
per pixel for ARCONS at Palomar, from measurements of the pixel
separation of known visual binaries) together completely determine the
astrometric solution. A list of photon-event pixel coordinates can
then be transformed into RA and Dec coordinates using linear matrix
translation and rotation operations.

This procedure assumes that there is negligible field distortion on
the scale of the science camera FOV. In fact there is known to be a
small amount of distortion across the field, perhaps up to
$\sim 0.5\farcs$, along with a certain amount of chromatic
aberration; this remains to be fully characterized, however. In principle
calibrations of the larger guide-camera FOV, discussed in the next
section, could be used to remove this effect.

The images used for the astrometric solution are created by
integrating over a specified duration, chosen to be short enough that
field rotation is small ($<1$ pixel). This sets the sampling rate of
the astrometric solution time series. The default astrometric solution
time is 30 seconds. Faster times on brighter objects may allow
effective software tip/tilt corrections to improve the recovered
object PSF.

\subsubsection{Guide Camera Astrometry}

In order to find objects that are too faint to see on the science
camera in short exposures (i.e., shorter than the field rotation
timescale), we can use guide camera images to calculate the expected
position of the object on the ARCONS science array, since the location
of the ARCONS array is fixed with respect to the guide camera
detector. The guide camera used in ARCONS is a SBIG STF-8300M CCD
camera with a $1\farcm 5$ FOV. Pick off mirrors positioned to either
side of the ARCONS dewar entrance fold an image back to the guide
camera, which sees a wider field than ARCONS. An empty strip
down the center of the guider image marks the gap between the
mirrors, within which the ARCONS detector itself lies. Each guide
image is saved as a standard FITS file. Exposures of up to
$10\seconds$ are used to reveal stars down to $\sim$20th
magnitude. Knowing the catalog RA/Dec of the observed guide objects
along with parameters relating guide camera positions to ARCONS positions, it
is possible to assign RA/Dec coordinates to individual ARCONS pixels.

In practice, it is first necessary to measure and eliminate all field
distortions, which become significant on the scale of the guide camera
FOV. We use the Simple Imaging Polynomial image distortion convention
\citep[SIP; see][]{SIPConvention}, which uses a set of polynomials to
characterize the distortion. We use Source Extractor
\citep{SExtractor} to automatically locate objects in the guide
images; if the guider field is crowded or the objects are highly
distorted, the code also allows for manual extraction of objects
within the field. In both cases, we query the online VizieR
service\footnote{\url{http://vizier.u-strasbg.fr/viz-bin/VizieR}} for
object coordinates (currently primarily using the 2MASS and USNO-B1.0
catalogus, depending on source brightness). The objects in the
catalog then need to be matched with those found in the guider field.
To facilitate automatic processing, a list of potential match patterns
between the catalog stars and guider field stars is constructed. For
each potential match pattern, both a linear offset and a rotation are
applied to minimize the distance between catalog and image stars, and
the corresponding total of the squared residuals is recorded. The
pattern with least total error is used to associate objects found in
the guider field to those in the catalog. Manual matching is also an
option if the process fails.

In the FITS convention, any distortion parameters should not contain
zeroth or first order correction to an image. These two corrections
are included in the standard header keywords \textit{CD} and
\textit{CRVAL} (which describe offset, rotation, and linear
scaling). Linear offset and rotation are therefore first applied to
each guider image to eliminate lower order corrections. The distortion
polynomial is then calculated using multiple fields observed at
various times throughout an observing run, again by minimizing the sum
of the squared distances between matching stars in the catalog and the
guider field.  Unlike offset and rotation, the distortion parameters
are expected to hold approximately constant during a run, as the
optics setup remains unaltered during observations. (Since
  ARCONS is currently removed from the telescope between runs, alignment of
  the optical train is not, however, expected to be consistent from run
  to run.)

To construct a coordinate grid on the ARCONS array, we need to know
the location of at least one guide image pixel in the ARCONS pixel
coordinate frame. Since the ARCONS array itself cannot be seen in the
guider images, we manually determine the position of a known bright
star on the ARCONS array and compare it to the expected location in
the guide images, and from that obtain a reference point. Once this
fixed reference point is calibrated, and assuming the respective plate
scales of ARCONS and the guide camera are known, it is then possible
to assign RA/Dec coordinates to each ARCONS pixel, and extrapolate the
position of faint objects in the science array. In principle,
  knowledge of the field distortion established from the guide camera
  can also be used to mitigate the effects of distortion in the science
  camera astrometry; this has yet to be implemented, however.

At this point the technique of obtaining astrometric solutions for the
ARCONS science array using guide camera images remains somewhat
experimental, and has not yet been fully integrated into the standard
pipeline; currently, astrometry for long exposures and mosaics
(Section \ref{sec:imaging}) is registered using bright sources in the
science camera images.

\subsection{Calibrated Photon List Creation}\label{sec:photlist}
Once all calibrations have been applied, the data are written out as
fully calibrated photon lists, one HDF5 file per input raw file. Each
list is stored as a single large table, with one row per photon, and
columns recording various data about each photon, including: photon
arrival time relative to the start of the data file; original $x,y$
pixel location on the detector; right ascension and declination; hour
angle (used later for image reconstruction purposes); wavelength;
weights for the photon associated with flatfield and spectral response
calibrations; and any flags from the photon calibration
process. Unusable photons, such as those falling within hot or bad
intervals identified in the exposure time masks (Section
\ref{sec:timemasks}), or pixels with failed wavelength calibrations,
are omitted from the list in order to conserve data storage
space. (The storage space required increases in proportion to the
total flux recorded, so hot pixels in particular can have a
significant impact on total storage requirements.) Stored alongside
the photon list table in each photon-list file is a copy of each of
the calibration files used in creating the list, which can be used to
trace back and debug a given reduction. The relative size of these
calibration files is small enough that the resulting duplication of
calibration information across different photon list files does not
have a major impact.

\section{Applications of Calibrated Photon Lists}\label{sec:outputproducts}

The calibrated photon lists described in the previous section
constitute the basic data product from which science data products can
be derived. In this section, we show the use of these photon lists to
produce examples of imaging, time-series photometry, and spectroscopy
that show the multi-dimensional capabilities of MKIDs. In particular
we show results from commissioning observations of the Crab Pulsar, a
target which provides for all three applications using the same
dataset. We also show light curves of the eclipsing AM CVn binary SDSS
J0926+3624 from the same commissioning run. These demonstrate
multi-band time series photometry on timescales of order hours, in
comparison to the fast $33\,\mathrm{ms}$ period of the Crab Pulsar.

\subsection{Imaging}\label{sec:imaging}
\subsubsection{Image stacking}\label{sec:imagestacking}
The photon arrival time resolving capabilities of MKIDs can simplify observing
in many ways: the telescope can be dithered, nodded, offset, or even
repointed while the detector is exposing, since these can all be
accounted for in software after the fact. Guiding errors are of little
concern (unless it is required to maintain a image on the same pixel
at all times), and field rotation is likewise easily handled. Given
enough flux, even simple tip-tilt correction can in principle be
applied {\em ex post facto} to improve image quality, though this
capability remains to be developed in the pipeline.

For long exposures and any kind of
data stacking from separate observations, it is essential to perform
derotation and pointing corrections. An approach somewhat analogous to
the Variable-Pixel Linear Reconstruction (``Drizzle'') algorithm of
\citet{Fruchter2002} is used: flux from each detector pixel is
remapped (``drizzled'') onto a more finely sampled ``virtual pixel
grid'' after rotating and rescaling, accounting for the overlap areas
of the original and virtual pixels to conserve flux. In this case,
however, instead of drizzling flux (a continuous quantity) onto the
virtual grid, we are drizzling discrete photons.

Each photon in a calibrated photon list has an assigned RA and Dec
corresponding to the center location on the sky at the time of
detection for the detector pixel on which the photon was recorded. In
the absence of further knowledge of the responsivity across the
surface of the pixel, we assume that the detected photon could have
landed anywhere within its footprint with equal probability. Knowledge
of the hour angle at which the photon was observed allows us to
establish the orientation of the pixel on the sky. We can therefore
construct two unit vectors corresponding to the directions of the two
axes of the pixel on the sky, and add a uniform random offset
(positive or negative) along each of these unit vectors, so that the
photon is positioned randomly within the footprint of the detector
pixel in RA and Dec. This is repeated for all photons. A simple
2D histogram of the resulting RA, Dec coordinate pairs
then results in an image remapped onto a virtual grid in sky
coordinates, where the counts in each virtual pixel correspond to the
number of photons that have landed in that pixel. The individual
photon weights from flatfielding and spectral response rectification
are accounted for by weighting the histogram, so that instead of each
photon being considered a single count, it is multiplied by its
respective weights. Flux is naturally conserved, and is automatically
distributed across the virtual pixels in a way that accounts for their
overlap area with the detector pixels. Although we are drizzling
discrete photons, we note that applying the weights will still
generally yield non-integer counts, just as for conventional CCD image
flatfielding; attempts to round the data at any stage will only add
artificial quantization noise.

Since the ARCONS detector oversamples the PSF of the telescope image,
loss of image resolution as a result of the image reconstruction
process, as discussed by \citet{Fruchter2002}, is not a great
concern. However, it is straightforward in principle to emulate the
effective ``shrinkage'' of the detector pixels suggested by the same
authors to further mitigate resolution loss, by simply reducing the
range of the random offset added to each photon location. This has not
yet been tested in the ARCONS pipeline.

\subsubsection{Exposure Time Weighting}

As the ARCONS field of view rotates and shifts, the detector footprint
will rotate and shift on the virtual image plane: dead pixels and the
edges of the detector will drift across the virtual image, so that
some virtual pixels will see more total exposure time than others. To
produce a useful image, therefore, it is also necessary to weight by
the effective exposure time in each virtual pixel. To produce an
exposure weight map for the virtual image, we slice the original
exposure into short time steps exactly matching the time steps used
for calculating the photon astrometry. Exposure time for each time
step can then be thought of as a conserved quantity analogous to
flux. The effective exposure time for each detector pixel within the
time step is calculated based on the time masks described in Section
\ref{sec:timemasks}, so that periods when a pixel went ``bad'' are
also excluded. Detector pixels which are permanently bad are also
automatically excluded in this way.  These effective exposure times
are then distributed across the virtual output image grid in
proportion to the respective overlap areas of the detector pixels, in
the same way as flux is in the more conventional ``drizzle'' scheme,
where flux is handled as a continuous quantity (rather than the
individual photon events described above). The detector/virtual pixel overlap areas are
calculated using the Fortran routine {\tt boxer}, borrowed from the
``PyDrizzle'' package in the STSCI\_PYTHON distribution.\footnote{See
  \url{http://www.stsci.edu/institute/software_hardware/pyraf/stsci_python}}

\begin{figure*}
\epsscale{1.12}
\centering
\plotone{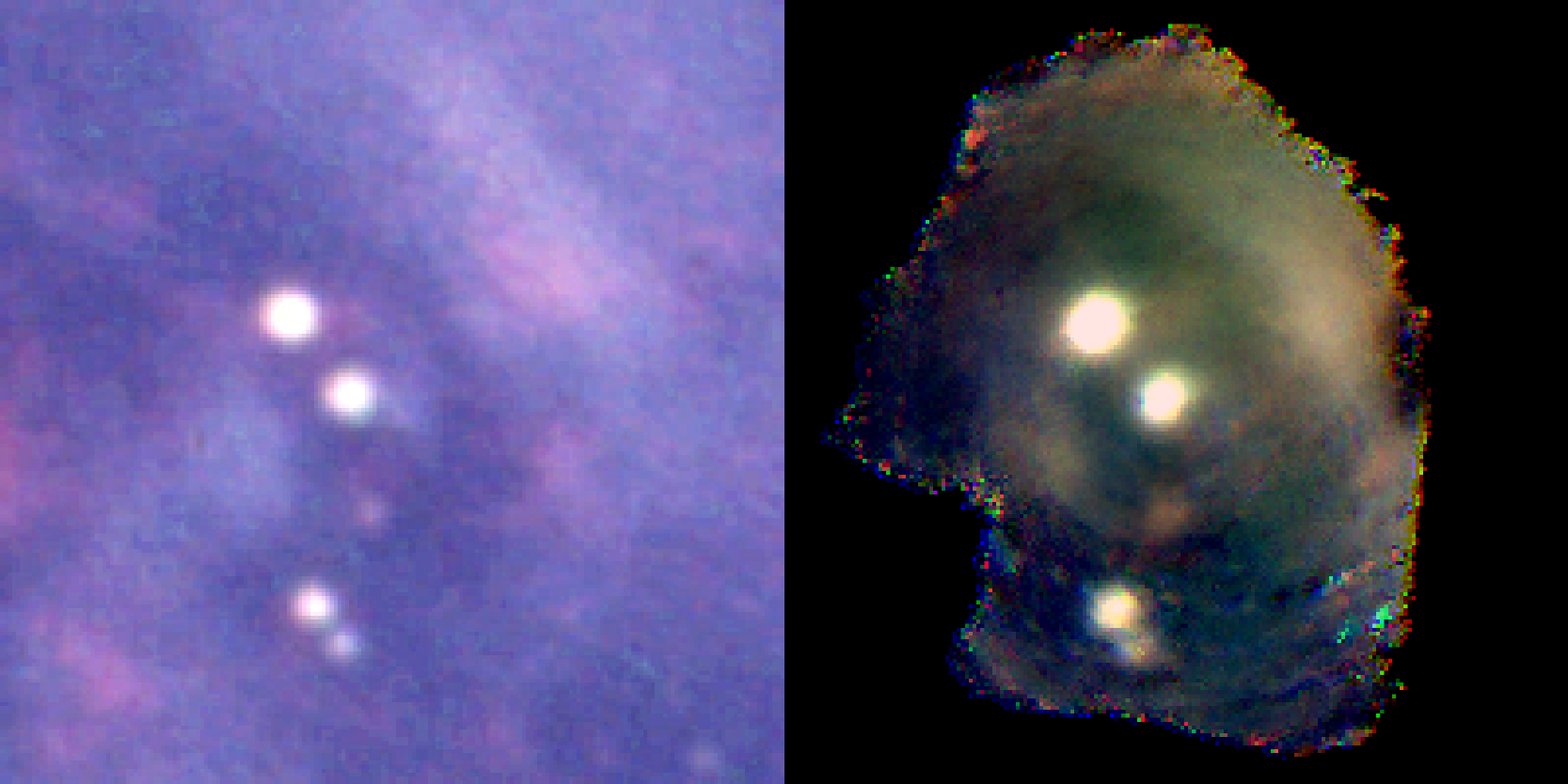}
\caption{\textit{Left}: Image of the Crab region taken with Ektachrome
  color film at the prime focus of the Kitt Peak 4-meter telescope
  (Credit: Bill Schoening/NOAO/AURA/NSF, 1973). \textit{Right}:
  stacking of $\approx 3\hr$ of exposure time on the Crab Pulsar taken
  with ARCONS in December 2012. RGB rendering is created from stacked
  photon lists in three wavebands: $4000-5000\ang$ (blue),
  $5000-6000\ang$ (green), and $6000-7000\ang$
  (red). The data are split into 5-minute chunks (matching the
  original observation files), which here are median combined to help
  minimize the effect of uncaught bad pixels. Streak-like artifacts
  in the image are due to remaining hot or otherwise badly behaved
  pixels that have not been handled by the pipeline and remain to be
  addressed. Higher noise around the image edges is caused by the low
  total effective exposure time in those areas. Since there is no
  atmospheric dispersion correction in the optics train, ARCONS
  resolves the differential refraction; the RGB component images have
  been reregistered by hand to reduce the effect, but some residual
  dispersion can still be seen.}
\label{fig:crabimage}
\end{figure*}

Multiplication of the virtual stacked image by the exposure weight map
(or equivalently, division by the per-pixel effective exposure times)
then gives an exposure weighted image which accounts fully for dead
areas, missing data, etc.
Figure \ref{fig:crabimage} shows a full image stack from $\approx 3
\hr$ of integration time on the Crab Pulsar, split into three
wavebands and rendered as an RGB image.

\subsection{Photometry}\label{sec:photometry}
The ARCONS pipeline provides photometry by both PSF fitting and
aperture photometry. For observations where high-cadence light curves
are desired, PSF fitting or aperture photometry can be performed on
data cubes generated directly from the observation files, still in
detector coordinates. For observations with long total integration
times, where field rotation is significant, or where dithering
  has been performed, we can perform standard aperture photometry on
the fully calibrated photon-list generated image in RA/Dec coordinates.

For most lightcurve analyses we are interested in short effective
integrations, meaning there is not sufficient time for field rotation
or dithering to fill in missing pixels. In these instances we can use
PSF fitting to estimate the missing flux. In its current state this
module fits a circular 2D Gaussian to the desired
wavelength slice in the data cube, still in detector
coordinates.  The baseline in this fit is subtracted off, removing the
sky background, and then the amplitude and width of the fitted
Gaussian are used to calculate the flux from the object.

\begin{figure}
\epsscale{1.12}
\centering
\plotone{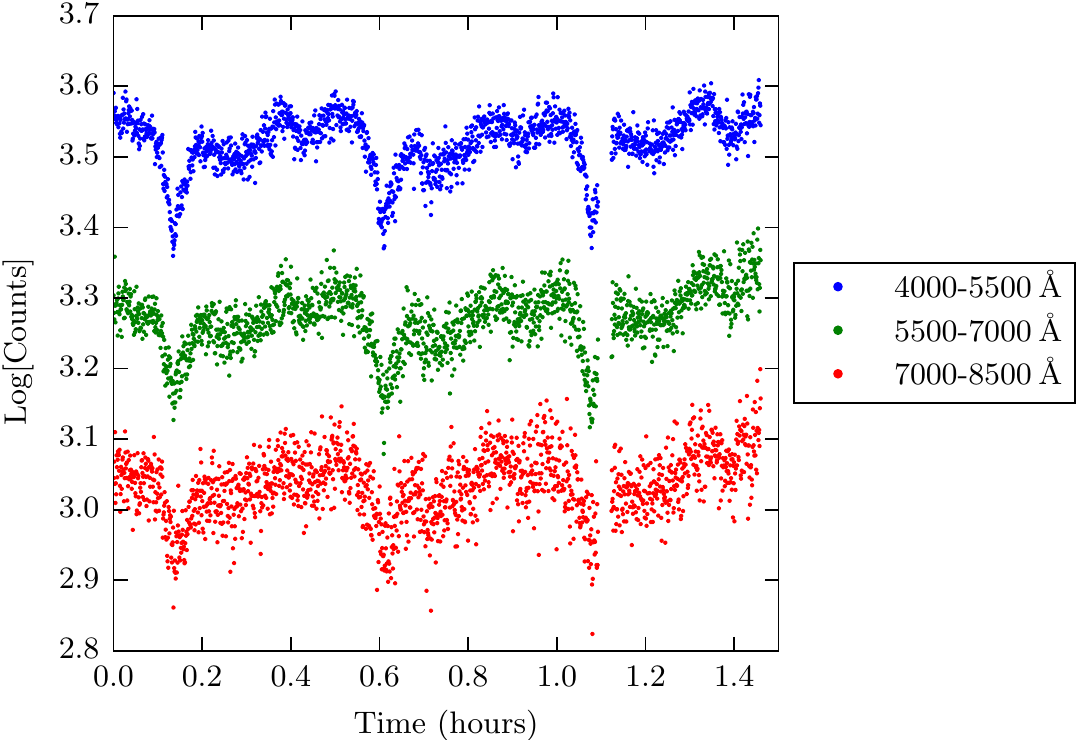}
\caption{A set of light curves generated using simple Gaussian
    PSF fitting photometry on partially calibrated data in detector
    coordinate space, demonstrating the multi-wavelength capability of
    ARCONS. The photometry shows the eclipsing AM CVn star SDSS
    J0926+3624 in blue, green, and red bands (from top to
    bottom). Average standard deviations over a flat part of the
    lightcurve on short timescales are 118, 89, and 77 counts per
    $3\sec$ integration, for the three wavelength bands,
    respectively. By comparison, the propagated PSF fitting errors are
    138, 113, and 103 counts (larger likely because of systematic
    mismatch between the assumed Gaussian PSF model and the actual
    instrumental PSF). The shapes of the light curves are consistent
    with expectations, but we found no chromatic dependency, and the
    light curves match across the bands within our precision
    \citep[see][]{Szypryt2014}.}
\label{fig:lightcurve-multiband}
\end{figure}

When observing bright targets, such as planet transits, we
must defocus the telescope to prevent count-rate saturation,
often by a substantial amount.  We are currently developing a robust
module to perform PSF fitting on non-Gaussian PSFs to analyze such
observations. This development is ongoing and is not included in the
current pipeline release. An example series of multiband light
  curves from the eclipsing AM CVn star, SDSS J0926+3624, created
using our PSF fitting pipeline, is shown in
Figure~\ref{fig:lightcurve-multiband} \citep[for more detail,
  see][]{Szypryt2014}.

For applications where PSF fitting fails (e.g., due to high sky
background when analyzing far-red slices of our data cubes), we can apply
standard aperture photometry to the data. This method is only applicable if the telescope pointing is
stable enough to maintain the object's position on the same pixel over
the course of the observation. Otherwise the PSF will drift over a
changing pattern of missing pixels and PSF fitting is required to
estimate the varying amount of missing flux. In our experience the
Hale $200\inch$ telescope at Palomar is stable enough for us to maintain our
target position to approximately one of ARCONS'
$\approx0\farcs 44$ pixels.

\begin{figure}
\epsscale{1.12}
\plotone{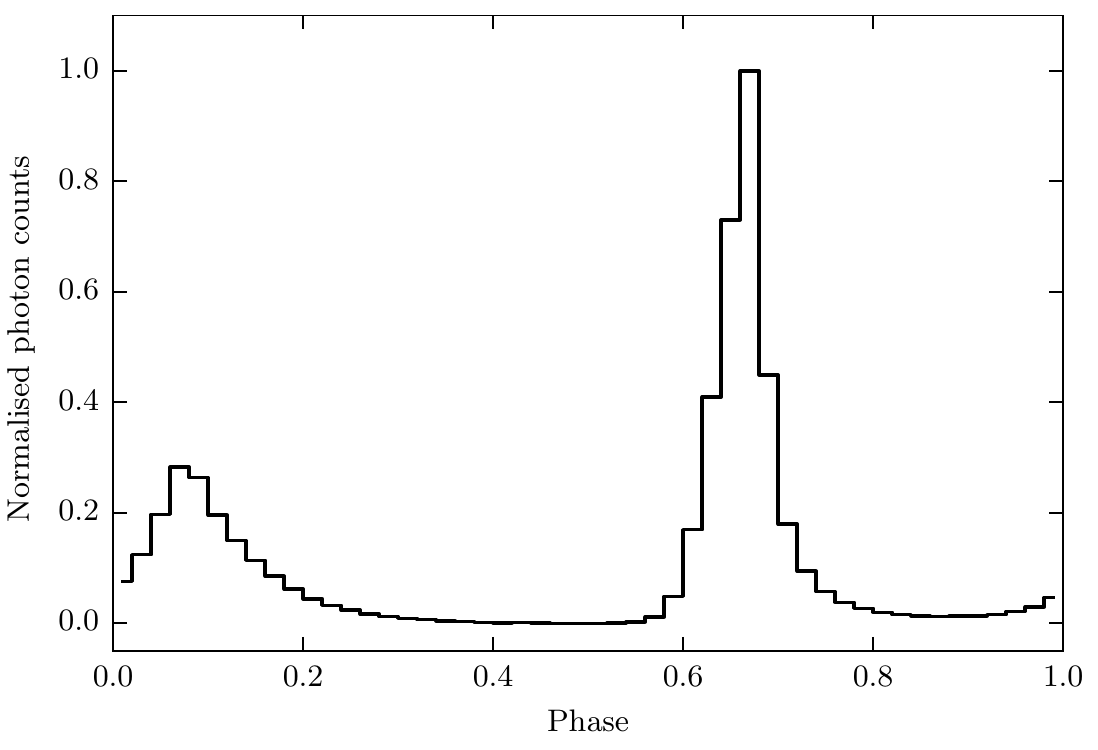}
\caption{Pulse profile of the Crab Pulsar from aperture photometry on
  a four-dimensional ``data hypercube'' generated from fully
  calibrated photon lists in RA/Dec space, demonstrating
    ARCONS' high time resolution. Data are folded on the pulsar's
    $33\,\textrm{ms}$ period to build adequate signal. The off-pulse level is
  taken to represent the background sky/nebulosity level, and is
  subtracted. Owing to the high integrated photon count, error
    bars are too small to be visible in this plot.}
\label{fig:crabProfile}
\end{figure}

For observations where short exposure times are not required
(or not feasible), we can perform aperture photometry on the
fully calibrated photon list images in RA/dec space, where the
  astrometry for each photon has been fully determined. Figure
\ref{fig:crabProfile} shows the broadband optical pulse profile of the
Crab Pulsar derived from performing aperture photometry on the
calibrated photon lists that produced Figure
\ref{fig:crabimage}. In this case the timestamps of all
  photons in the aperture were folded on the pulsar period using an
  ephemeris derived from the simultaneous radio data (see Section
  \ref{sec:barycenter}), in order to build up sufficient signal. The
  dithering and field-rotation during the observations mean that
  missing pixels are not a concern.  More details can be found in
\citet{Strader2013}.

\subsection{Spectroscopy}\label{sec:spectroscopy}
\begin{figure}
\epsscale{1.12}
\centering
\plotone{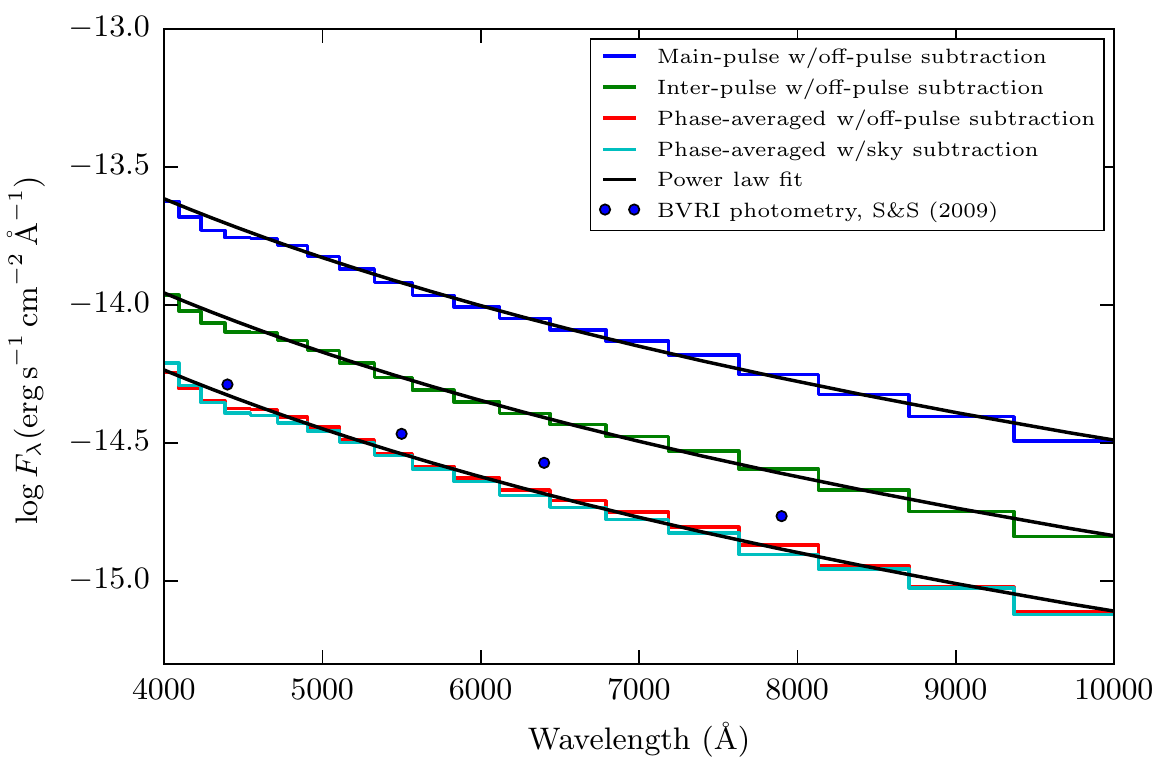}
\caption{Spectra of the Crab Pulsar during the main pulse, inter
  pulse, and phase-averaged over the whole period, demonstrating the
  ability of ARCONS to perform time-resolved spectroscopy. All spectra
  are corrected for the effects of interstellar reddening according to
  $A_{\lambda}/E_{B-V} = -7.51\log(\lambda/\lambda_0) + 1.15,$ with
  $\lambda_0 = 10000\ang$ and $E_{B-V} = 0.51$
  \citep{Carraminana2000}. The phase-averaged spectra are integrated
  over all time, including the off-pulse time, resulting in a lower
  mean flux than the main- and inter-pulse spectra. The phase averaged
  spectrum generated by using the sky annulus subtraction is almost
  identical to that generated by using the off-pulse aperture
  subtraction. BVRI photometry points taken from \citet{Sandberg2009}
  are also de-reddened and overplotted for comparison. The
  phase-averaged spectra follow the same trend as the photometry for
  all points, with the spectra falling within $\approx 10$\% of the
  BVRI photometry. This difference likely results from variations in
  flux from the Crab Pulsar and its surrounding nebula
  \citep{Pacini1971, Nasuti1996, Sandberg2009} and potential
  calibration errors for the ARCONS MKIDs array. The spectra of the
  main-pulse and inter-pulse are generated using the off-pulse
  aperture subtraction. Fits for the various spectra were found using
  a power law, $F_{\lambda}=K(\lambda/\lambda_0)^{-\alpha-2},$ with
  $\lambda_0=6000\ang$ \citep[per][]{Carraminana2000}, and
  $4808\ang<\lambda<9012\ang$. The main-pulse and inter-pulse were
  found to have spectral indices of $\alpha_{\lambda} = +0.20$ and
  $\alpha_{\lambda} = +0.21$ respectively, agreeing with the values
  reported by \citet{Carraminana2000} $(\alpha = +0.2\pm0.1$ for both
  pulses). The spectral index of the phase averaged spectrum was found
  to be $\alpha_{\lambda} = 0.20$. Note that flux error
    estimates are too small to be appear in the plot (typically
    $\sim 10^{-4}$ for our data, and $\sim 10^{-3}$ for the BVRI
    points from \citet{Sandberg2009}, in log space.)}
\label{fig:crabspec}
\end{figure}

Spectroscopy can be performed on a multi-dimensional image
stack. During the photon stacking process, there is an option to
include a wavelength dimension along with the RA and Dec positions.
This is done by additionally binning the photons by wavelength, making
a three-dimensional histogram of the RA, Dec, and wavelengths. The
result is a data cube that can be summed across the spectral dimension
to produce an image remapped onto a virtual grid in sky coordinates as
discussed in Section \ref{sec:imaging}, or summed across chosen areas
in RA and Dec to obtain a spectral energy distribution. Spectra are
found using aperture spectroscopy, similar to standard aperture
photometry, but with the photon counts per pixel binned by their
respective wavelengths. The resulting total photon counts per
wavelength of the object can then be converted to flux and plotted
against wavelength to produce a spectrum. Figure \ref{fig:crabspec}
shows the results of aperture spectroscopy for a Crab Pulsar data
cube.

By including an additional dimension in the histogram, it is also
possible to include information about photon arrival time, making a
four-dimensional histogram against RA, Dec, wavelength, and time. This
four-dimensional data ``hypercube'' can be summed across any of its
dimensions as desired, and sliced in order to obtain resolution in any
dimension. This can be used to obtain the various products already
described; it can also be used to obtain time-resolved spectra, for
example.  In the case of the Crab Pulsar, the emissions blink on and
off very rapidly, with a period of $33\,\textrm{ms}$. Figure
\ref{fig:crabProfile} clearly shows that there are two main
contributors to the overall flux coming from the object: the large
main pulse, and the smaller inter-pulse. Since the Crab Pulsar is
highly periodic, it is possible to bin the time axis by the pulsar's
phase. After preparing a ``hypercube'' with dimensions of RA, Dec,
wavelength, and pulsar phase, we can select photons where the pulsar
is ``off'' and use that to make an estimate of the sky background
spectrum \textit{within} the photometric aperture itself, rather than
using a sky annulus outside the aperture as an approximation, where
nebulosity may affect the results. Similarly, it is possible to
determine the spectrum for the main pulse and inter-pulse
separately. The spectrum of the off-pulse can then be subtracted from
the main-pulse and inter-pulse spectra at that same position for every
virtual pixel within the aperture. The result is the sky-corrected
spectrum of the main pulse and inter pulse.  Figure \ref{fig:crabspec}
shows these phase-resolved spectra for the Crab Pulsar, showing good
agreement with previously published data
\citep{Carraminana2000,Sandberg2009}. This technique demonstrates the
full multi-dimensional capabilities of ARCONS, using imaging
information to produce wavelength-resolved spectra with very fast time
resolution.

\section{Summary}\label{sec:conclusions}

The pipeline as it currently stands takes the initial photon event
detections from the instrument firmware, and passes them through the
following main components: data calibration (cosmic-ray masking,
bad-pixel masking, wavelength calibration, flatfielding, absolute
spectral response calibration, astrometry, nonlinearity correction);
calibrated photon list creation; and three output products, imaging,
time-series photometry, and spectroscopy. The output products may all
be based either on calibrated photon lists or on calibrated versions
of the raw data files, depending on requirements and on how much field
rotation and/or dithering or repointing has occurred over the total
exposure time of interest.

Not all of these components are fully developed, and we expect much
room for improvement as we begin to understand better the
idiosyncrasies of the technology. Processing speeds are currently
quite slow (on the order of a few times real-time), and as we move to larger
array formats and increased observing time, performance will become an
increasing concern. Translating time-consuming components to C should
help, since native Python is not particularly fast. The massive
datasets involved also lend themselves well to parallelization, which
should yield significant speed gains. Such parallelization will be
critical for high-speed applications such as speckle nulling
\citep[][S.~R.~Meeker et al., 2015 in preparation]{Martinache2014},
where interference speckle noise in high-contrast coronagraphic images
must be tracked on atmospheric seeing timescales ($\sim
10^{-2}\seconds$).

Four-dimensional image ``hypercubes'' (with dimensions of RA, Dec,
wavelength, and time) remain to be fully implemented as an easily
manageable final output product, though the elements to do so are
largely in place. Other avenues for improvement to the pipeline
include using the full wavelength- and time-resolved information
available in the data to better characterize and discriminate hot
pixels (to the extent that this is necessary with future generations
of MKID arrays); fully integrating astrometric solutions from guide
camera data rather than science camera data (to facilitate observing
very faint objects); performing \textit{ex post facto} image tip-tilt
corrections for bright objects; improving our handling of
photon-pileup and better understanding nonlinearity effects; and
improved integration, streamlining, and automation of the pipeline.

MKID technology is still relatively young and rapidly developing.  Our
data reduction pipeline is therefore necessarily a work in progress,
and is written with an eye to being readily adaptable to new and
upgraded MKID arrays. The current ARCONS pipeline development code is
made available at \url{https://github.com/bmazin/ARCONS-pipeline}.

%------------------------------------------------------------------
\acknowledgments
We thank Shri Kulkarni and Tom Prince for the use of the Hale
Telescope, and the staff of Palomar Observatory for their generous
support. We are grateful to our anonymous referee for a careful and
helpful review of this work. The MKID detectors used in this work were
developed under NASA grant NNX11AD55G. The MKID digital readout was
partially developed under NASA grant NNX10AF58G. SRM and PS are
supported by NASA Office of the Chief Technologist’s Space Technology
Research Fellowships (NSTRF). JCvE was funded by a National Science
Foundation ATI grant, number AST-1308556. This work was partially
supported by the Keck Institute for Space Studies. Fermilab is
operated by Fermi Research Alliance, LLC under Contract
No. De-AC02-07CH11359 with the United States Department of
Energy. STSCI\_PYTHON is a product of the Space Telescope Science
Institute, which is operated by AURA for NASA. This research made use
of Astropy, a community-developed core Python package for Astronomy
\citep{astropy}. Based on observations obtained at the Hale Telescope,
Palomar Observatory as part of a continuing collaboration between the
California Institute of Technology, NASA/JPL, NOAO, Oxford University,
Stony Brook University, and the National Astronomical Observatories of
China.

{\it Facility:} \facility{Hale (ARCONS)}

%------------------------------------------------------------------

% Make List of References (BibTeX implemented)
\bibliography{apj-jour,arconsbibliography}

\end{document}